\documentclass{ptptex}

\usepackage{amsmath}
\usepackage{graphicx}
\usepackage{wrapft}

\newcommand{\bsigma}{\mbox{\boldmath $\sigma$}}

\newcommand{\bpi}{\mbox{\boldmath $\pi$}}
\def\nn{\nonumber}

\title{Pseudospin and Deformation-induced Gauge Field in Graphene}

 \author{
Ken-ichi SASAKI,$^{}$\thanks{E-mail: sasaken@flex.phys.tohoku.ac.jp}
and Riichiro SAITO
}

\inst{
Department of Physics, Tohoku University,
Sendai 980-8578, Japan 
}

\abst{
 The basic properties of $\pi$-electrons near the Fermi level
 in graphene are reviewed from a point of view of the pseudospin
 and a gauge field coupling to the pseudospin.
 The applications of the gauge field 
 to the electron-phonon interaction and to the edge states 
 are reported.
}

\begin{document}
\maketitle

\section{Introduction}\label{sec:intro}

The electronic properties of a single layer of graphite,
graphene,~\cite{novoselov05,zhang05,heersche07} 
have attracted much attention
due to the ``relativistic'' character 
of $\pi$-electrons near the Fermi level.
The energy band structure of graphene exhibits
a linear energy dispersion relation 
around the two inequivalent, 
hexagonal corners of the first Brillouin zone
in the $k$-space 
(the K and K' points).~\cite{wallace47,slonczewski58}
The wavefunction (Hamiltonian) of $\pi$-electrons 
has two component (a $2\times 2$ matrix) form
due to the fact that 
the unit cell of graphene consists of two carbon atoms
(A and B atoms).
The effective-mass Hamiltonian of $\pi$-electrons 
around the K point or the K' point
is given by linear momentum operator,
which is relevant to the linear energy dispersion relation
of graphene.
The effective-mass equation is similar to 
the massless Dirac equation or the Weyl equation.~\cite{sakurai67}

The original Dirac equation for an electron has the form
\begin{align}
 i\hbar \frac{\partial}{\partial t}\psi({\bf r},t) 
 = 
 \begin{pmatrix}
  -c \bsigma \cdot {\hat {\bf p}} & I m c^2 \cr
  I m c^2 & c \bsigma \cdot {\hat {\bf p}} 
 \end{pmatrix}
 \psi({\bf r},t),
 \label{eq:dirac}
\end{align}
where $c$ is the speed of light,
$m$ is the mass of electron,
and ${\hat {\bf p}}$ $(=-i\hbar \nabla)$ is the momentum operator.
The Hamiltonian is a $4\times 4$ matrix, 
which is written in terms of the Pauli matrices
$\bsigma=(\sigma_x,\sigma_y,\sigma_z)$ and
$2\times 2$ identity matrix $I$.~\footnote{
We use the Pauli matrices of the form of
$\sigma_x
=\begin{pmatrix}
  0 & 1 \cr 1 & 0
 \end{pmatrix}$, 
$\sigma_y
=\begin{pmatrix}
  0 & -i \cr i & 0
 \end{pmatrix}$, and
$\sigma_z
=\begin{pmatrix}
  1 & 0 \cr 0 & -1
 \end{pmatrix}$.
The $2 \times 2$ identity matrix $I$ is given by
$I = \begin{pmatrix}
 1 & 0 \cr 0 & 1
\end{pmatrix}$.
}
$\psi({\bf r},t)$ of the Dirac equation is 
a four-component wavefunction,
which naturally explains 
the spin degree of freedom for particle (electron)
and the antiparticle (positron).
In the massless limit ($m=0$), 
the Dirac equation is split into two 
equations of a $2\times 2$ matrix form, that is,
the Weyl equations for massless neutrinos.
Setting $\psi={}^t(\psi_L,\psi_R)$ in Eq.~(\ref{eq:dirac}), 
we obtain the Weyl equation for $\psi_R$ and $\psi_L$ as
\begin{align}
\begin{split}
 & i\hbar \frac{\partial}{\partial t}\psi_L({\bf r},t) 
 = -c \bsigma \cdot {\hat {\bf p}} \psi_L({\bf r},t), \\
 & i\hbar \frac{\partial}{\partial t}\psi_R({\bf r},t) 
 = c \bsigma \cdot {\hat {\bf p}} \psi_R({\bf r},t).
\end{split}
 \label{eq:weyl}
\end{align}
The two component of each Weyl equation
represents the spin degree of freedom.

On the other hand,
the effective-mass equations for the K and K' points
of graphene are written as
\begin{align}
 \begin{split}
  & i\hbar \frac{\partial}{\partial t}\psi^{\rm K}({\bf r},t) 
  = v_{\rm F} \bsigma \cdot {\hat {\bf p}} \psi^{\rm K}({\bf r},t), \\
  & i\hbar \frac{\partial}{\partial t}\psi^{\rm K'}({\bf r},t) 
  = v_{\rm F} \bsigma' \cdot {\hat {\bf p}} \psi^{\rm K'}({\bf r},t),
 \end{split}
 \label{eq:weyl_graphene}
\end{align}
where $v_{\rm F}$ is the Fermi velocity,
$\bsigma=(\sigma_x,\sigma_y)$,
$\bsigma'=(-\sigma_x,\sigma_y)$,
and ${\hat {\bf p}}=({\hat p}_x,{\hat p}_y)$.
The equation for $\psi^{\rm K}$ of Eq.~(\ref{eq:weyl_graphene}) 
is similar to the Weyl equation for $\psi_R$
of Eq.~(\ref{eq:weyl}) with $p_z=0$ and
by substituting $c$ to $v_{\rm F}$.
The equation for $\psi^{\rm K'}$ 
is similar to the Weyl equation for $\psi_L$ with $p_z=0$, 
the substitution of $c$ to $v_{\rm F}$, and 
the negative sign in front of $\sigma_y$.
Although the character of the two component of 
the effective-mass equations of graphene (A and B atoms) 
and of the Weyl equation (up spin and down spin)
is different from each other, 
the equation and the resulting solution for 
given $\sigma_x$ and $\sigma_y$ 
are the same.
Thus, it is appropriate that 
the two component structures 
$\psi_{\rm K}={}^t(\psi_{\rm K}^A,\psi_{\rm K}^B)$
and $\psi_{\rm K'}={}^t(\psi_{\rm K'}^A,\psi_{\rm K'}^B)$
of the effective-mass equations
for graphene near the Fermi level 
is referred to as the pseudospin.
A pseudospin structure gives 
a rich variety of interesting physical phenomena
of graphene and nanotubes.
It is known that 
absence of a backward scattering 
of an electron in graphene and carbon nanotubes
is relevant to the nature of pseudospin,~\cite{ando98,ando05}
in which a $2\pi$ rotation of a pseudospin wavefunction
around the K point in the two-dimensional Brillouin zone 
does not gives the original wavefunction 
but gives minus sign to the wavefunction.
The pseudospin in the $k$-space behaves
similarly to the real spin in the real space.

The interaction between 
an electron and an electromagnetic field 
is given by replacing ${\hat {\bf p}}$ in the Dirac equation
with the kinematical momentum 
${\hat \bpi} = {\hat {\bf p}}-e{\bf A}({\bf r})$
where $-e$ is the charge of electron
and ${\bf A}({\bf r})$ is a vector potential.
The spin of an electron is polarized by a magnetic field,
${\bf B}({\bf r})= \nabla \times {\bf A}({\bf r})$,
which can be shown explicitly by the Dirac equation.~\cite{sakurai67}
In the non-relativistic limit,
the Dirac equation reduces to the Pauli equation
in which the leading interaction 
between spin and a magnetic field
is reduced to the Zeeman term:
\begin{align}
 -\frac{e\hbar}{2m} \bsigma\cdot {\bf B}({\bf r}).
 \label{eq:zeeman}
\end{align}
Since the equation for graphene 
is similar to the Dirac equation,
the following question arises;
What is the field that polarizes the pseudospin?
A magnetic field is a candidate.
The same procedure as that in Eq.~(\ref{eq:zeeman})
shows, however, that 
the pseudospin is not polarized by a magnetic field
because the Zeeman term appears with the opposite sign
at the K point and at the K' point as
\begin{align}
 \begin{split}
  & -\frac{e\hbar}{2m} \bsigma\cdot {\bf B}({\bf r}) \ \ ({\rm K}\ {\rm
  point}), \\
  & +\frac{e\hbar}{2m} \bsigma\cdot {\bf B}({\bf r}) \ \ ({\rm K'}\ {\rm
  point}).
 \end{split}
\end{align}
Thus the direction of the pseudospin polarization 
which is induced at the K point by a magnetic field
is opposite to that at the K' point by the same magnetic field.
Thus the pseudospin of graphene is not polarized 
by ${\bf B}({\bf r})$.~\footnote{
It is noted that even in the massless limit
of Eq.~(\ref{eq:zeeman}), 
we obtain a similar coupling term by taking
the square of the Hamiltonian in the Weyl equation as
\begin{align}
 \left(c \bsigma \cdot {\hat \bpi} \right)^2 
 = c^2 I \left({\hat {\bf p}}- e{\bf A}({\bf r})\right)^2 
 - e \hbar c^2 \bsigma \cdot {\bf B}({\bf r}).
 \label{eq:square_weyl}
\end{align} 
The first term in the right-hand side of Eq.~(\ref{eq:square_weyl})
does not concern the spin, but the second term 
in the right-hand side of Eq.~(\ref{eq:square_weyl})
is similar to the Zeeman term which shows that
the spin is polarized by a magnetic field.
\begin{align}
 \begin{split}
  &  \left(v_{\rm F} \bsigma \cdot 
  {\hat \bpi} \right)^2 
 = v_{\rm F}^2 I \left({\hat {\bf p}}- e{\bf A}({\bf r})\right)^2 
  - e \hbar v_{\rm F}^2 \sigma_z \cdot B_z({\bf r}), \ \ ({\rm K} \ {\rm
  point})\\
  &  \left(v_{\rm F} \bsigma' \cdot 
  {\hat \bpi} \right)^2 
 = v_{\rm F}^2 I \left({\hat {\bf p}}- e{\bf A}({\bf r})\right)^2 
  + e \hbar v_{\rm F}^2 \sigma_z \cdot B_z({\bf r}). \ \ ({\rm K} \ {\rm
  point})
 \end{split}
 \label{eq:square_weyl_2}
\end{align}
The opposite sign in front of $e \hbar v_{\rm F}^2 \sigma_z \cdot
B_z({\bf r})$ at the K point and at the K' point shows that
the pseudospin is not polarized by the magnetic field.
}
Mathematically, this observation leads us to assume
a new gauge field ${\bf A}^{\rm q}({\bf r})$
which has the opposite sign of ${\bf A}^{\rm q}({\bf r})$ 
at the K point to the K' point as
\begin{align}
 \begin{split}
  & i\hbar \frac{\partial}{\partial t}\psi_{\rm K}({\bf r},t) 
  = v_{\rm F} \bsigma \cdot \left({\hat {\bf p}}+{\bf A}^{\rm q}({\bf
  r}) \right) \psi_{\rm K}({\bf r},t), \\
  & i\hbar \frac{\partial}{\partial t}\psi_{\rm K'}({\bf r},t) 
  = v_{\rm F} \bsigma' \cdot \left( {\hat {\bf p}}-{\bf A}^{\rm q}({\bf
  r})\right)\psi_{\rm K'}({\bf r},t).
 \end{split}
 \label{eq:weyl_graphene_A}
\end{align}
Then the corresponding field defined by
${\bf B}^{\rm q}({\bf r})= \nabla \times {\bf A}^{\rm q}({\bf r})$
can polarize the pseudospin because the Zeeman term
appears as the same sign for the K and K' points 
in this case.

There is an example of the pseudospin-polarized state,
that is, a localized state appearing near the zigzag edge of graphene, 
which is called the edge states.~\cite{fujita96}
The wavefunction of the edge state
has a value only on A atoms when the zigzag edge atoms
consist of A atoms.
Thus we expect that a field ${\bf B}^{\rm q}({\bf r})$
appears around the edge.
We will show that the zigzag edge structure is relevant to
a field ${\bf B}^{\rm q}({\bf r})$ 
which polarizes the pseudospin
near the zigzag edge.~\cite{sasaki06jpsj}
A gauge field ${\bf A}^{\rm q}({\bf r})$ and 
a field ${\bf B}^{\rm q}({\bf r})$ is important 
not only for the localized edge states 
but also for the extended states.
For example, the electron-phonon (el-ph) interaction 
in graphene can be expressed by ${\bf A}^{\rm q}({\bf r})$,
which explains the chirality dependence 
of the el-ph interaction.~\cite{q1239}
In this paper we review our studies performed on 
the gauge field ${\bf A}^{\rm q}({\bf r})$ by
showing that 
the el-ph interaction and the edge boundary
are expressed by ${\bf A}^{\rm q}({\bf r})$.


Here, we would like to mention 
the relationship between our work and
previously published literature on 
deformation-induced gauge field.
Kane and Mele introduced a {\it homogeneous} 
deformation-induced gauge field 
in the effective-mass theory.~\cite{kane97}
The gauge field represents uniform lattice deformations 
such as uniform bend, twist and curvature of a carbon nanotube.
The uniform gauge field for the curvature of a nanotube
changes the boundary condition 
around a tube axis which is given by a generalized 
Aharanov-Bohm (AB) effect and 
induces a small energy gap in 
a chiral {\it metallic} carbon nanotube.
The curvature-induced mini gap was observed by scanning tunneling
spectroscopy (STS) experiment by 
Ouyang {\it et al}.~\cite{ouyang01}
A generalization of the gauge field to a
local field is necessary for describing a local lattice deformation.
In the previous paper,~\cite{sasaki05}
we generalized the gauge field introduced by Kane and Mele,
to include a local lattice deformation of graphene.
Further, by defining a deformation-induced magnetic field, 
we explained 
the {\it local} modulation of the energy band gap,~\cite{sasaki05} 
which was observed in the STS
measurement for pea-pod (C$_60$ encapsulated carbon nanotube) 
by Lee {\it et al}.~\cite{lee02}

This paper is organized as follows.
In \S~\ref{sec:weyl}, 
we derive the effective-mass equations for graphene 
and define the pseudospin.
We show that the pseudospin determines the elastic
scattering amplitude of an electron by impurity potentials.
In \S~\ref{sec:weyl_deform},
we derive the effective-mass equation 
in the presence of the lattice deformation,
in which gauge symmetry for ${\bf A}^{\rm q}({\bf r})$
is essential to the time-reversal properties of the el-ph interaction.
In \S~\ref{sec:elph},
the formulation explained in \S~\ref{sec:weyl_deform}
is applied to the el-ph interaction
for optical and acoustic phonon modes.
In \S~\ref{sec:edge},
we will discuss the edge state by ${\bf A}^{\rm q}({\bf r})$.
In \S~\ref{sec:dis},
summary of this paper is given.

\section{Effective-mass Theory}\label{sec:weyl}

\begin{figure}[htbp]
 \begin{center}
  \includegraphics[scale=0.7]{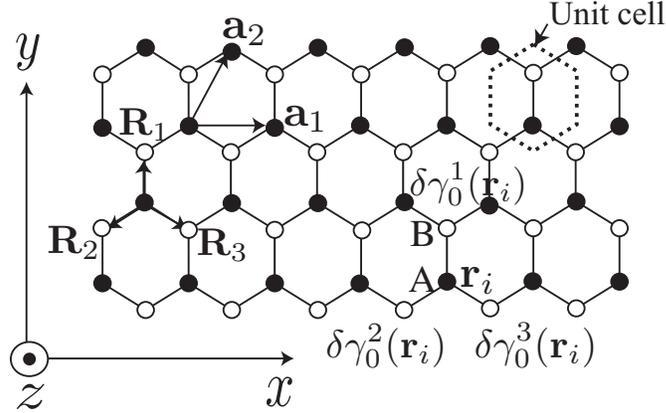}
 \end{center}
 \caption{
 A hexagonal unit cell of graphene consists of 
 {\rm A} (solid circle) and {\rm B} (open circle) atoms. 
 ${\bf a}_1$ and ${\bf a}_2$ are the unit vectors 
 ($a$ denotes the lattice constant: 
 $a\equiv|{\bf a}_i|=\sqrt{3}a_{\rm cc}$
 where $a_{\rm cc}$ is the C-C bond length.). 
 ${\bf R}_a$ ($a=1,2,3$)
 are vectors pointing to the nearest-neighbor  
 {\rm B} atoms from an {\rm A} atom.
 For the $xyz$ coordinate system,
 ${\bf R}_a$ are written as
 ${\bf R}_1=a_{\rm cc}{\bf e}_y$,
 ${\bf R}_2=-(\sqrt{3}/2)a_{\rm cc}{\bf e}_x -(1/2)a_{\rm cc}{\bf e}_y$,
 and 
 ${\bf R}_3=(\sqrt{3}/2)a_{\rm cc}{\bf e}_x -(1/2)a_{\rm cc}{\bf e}_y$
 where ${\bf e}_x$ (${\bf e}_y$)
 is the dimensionless unit vector for the $x$-axis ($y$-axis).
 Local modulations of the hopping integral are defined by 
 $\delta \gamma^a_0({\bf r}_i)$ ($a=1,2,3$) where ${\bf r}_i$ 
 ($i\in {\rm A}$) is the position of an A-atom.
 }
 \label{fig:graphene}
\end{figure}

In this section,
we derive an effective-mass Hamiltonian 
from the nearest-neighbor tight-binding model 
for a graphene.
The nearest-neighbor tight-binding model 
is given by
\begin{align}
 {\cal H}_0 = 
 -\gamma_0 \sum_{i \in {\rm A}} \sum_{a=1,2,3} 
 \left(
 (c_{i+a}^{\rm B})^\dagger c_i^{\rm A} + 
 (c_i^{\rm A})^\dagger c_{i+a}^{\rm B}
 \right),
 \label{eq:H0}
\end{align}
where $\gamma_0$ ($\approx 2.7$ eV) 
is the nearest-neighbor hopping integral, 
$c_i^{\rm A}$ ($(c_i^{\rm A})^\dagger$)
is the annihilation (creation) operator of 
$\pi$-electron for A-atom at the position ${\bf r}_i$,
and $c^{\rm B}_{i+a}$ ($(c^{\rm B}_{i+a})^\dagger$) 
is that for B-atom at ${\bf r}_{i+a}$ $(\equiv{\bf r}_i+{\bf R}_a)$
where ${\bf R}_a$ ($a=1,2,3$) are vectors 
pointing to the three nearest-neighbor B-atoms from an A-atom
(see Fig.~\ref{fig:graphene}).

We use the Bloch theorem to diagonalize Eq.~(\ref{eq:H0}).
The Bloch wavefunction with wavevector ${\bf k}$ 
is defined by
\begin{align}
 |\Psi_s^{\bf k} \rangle = 
 \frac{1}{\sqrt{N_u}} \sum_{i \in s} 
 e^{i {\bf k} \cdot {\bf r}_i} (c^s_i)^\dagger |0 \rangle \ \ \
 (s = {\rm A},{\rm B}),
 \label{eq:Bloch}
\end{align}
where the sum on $i$ is taken over the crystal, 
$N_u$ is the number of the hexagonal unit cells,
and $|0 \rangle$ denotes the state of carbon atoms 
without $\pi$-electrons.
The off-diagonal matrix element of ${\cal H}_0$ 
is given by
\begin{align}
 \begin{split}
  & \langle \Psi_{\rm A}^{\bf k} |{\cal H}_0| 
  \Psi_{\rm B}^{\bf k} \rangle
  = - \gamma_0 \sum_{a=1,2,3} f_a({\bf k})
  = - \gamma_0 f({\bf k}), \\
  & \langle \Psi_{\rm B}^{\bf k} |{\cal H}_0|
  \Psi_{\rm A}^{\bf k} \rangle
  = - \gamma_0 \sum_{a=1,2,3} f_a({\bf k})^*
  = - \gamma_0 f({\bf k})^*,
 \end{split}
 \label{eq:H0_bb}
\end{align}
where $f_a({\bf k})\equiv e^{i{\bf k} \cdot {\bf R}_a}$
and $f({\bf k})$ is given by~\cite{saito98book} 
\begin{align}
 f({\bf k})\equiv \sum_{a=1,2,3} f_a({\bf k}).
\end{align}
The diagonal matrix element of ${\cal H}_0$,
$\langle \Psi_{s}^{\bf k}|{\cal H}_0|\Psi_{s}^{\bf k}\rangle$,
($s=$A,B) can be taken to be zero.
The energy eigenequation is written 
by a $2\times 2$ matrix form as
\begin{align}
 E({\bf k})
 \begin{pmatrix}
  | \Psi_{\rm A}^{\bf k} \rangle \cr
  | \Psi_{\rm B}^{\bf k} \rangle
 \end{pmatrix}
 = -\gamma_0 
 \begin{pmatrix}
  0 & f({\bf k}) \cr
  f({\bf k})^* & 0 
 \end{pmatrix}
 \begin{pmatrix}
  | \Psi_{\rm A}^{\bf k} \rangle \cr
  | \Psi_{\rm B}^{\bf k} \rangle
 \end{pmatrix}.
 \label{eq:H_AB}
\end{align}
The energy band structure of graphene is obtained 
by solving 
\begin{align}
 \det
 \begin{pmatrix}
  E({\bf k}) & \gamma_0 f({\bf k}) \cr
  \gamma_0 f({\bf k})^* & E({\bf k})
 \end{pmatrix}
 =0,
\end{align}
whose solution is $E({\bf k})=+\gamma_0 |f({\bf k})|$,
($-\gamma_0 |f({\bf k})|$)
for the conduction (valence) energy band.
The conduction energy band and the valence energy band 
touch with each other at the K point,
${\bf k}_{\rm F}$ ($=(4\pi/3a,0)$),
and at the K' point, $-{\bf k}_{\rm F}$,
where $|f({\bf k})|$ vanishes.
By expanding $f_a({\bf k})$ in Eq.~(\ref{eq:H0_bb})
around the wavevector of ${\bf k}_{\rm F}$ (the K point),
we obtain 
\begin{align}
 f_a({\bf k}_{\rm F}+{\bf k})=f_a({\bf k}_{\rm F})
 +if_a({\bf k}_{\rm F}){\bf k}\cdot {\bf R}_a+\cdots.
 \label{eq:fa_KF}
\end{align}
Here ${\bf k}$ ($=(k_x,k_y)$) is measured from the K point.
Using ${\bf k}_{\rm F}=(4\pi/3a,0)$,
we get $f_1({\bf k}_{\rm F}) = 1$, 
$f_2({\bf k}_{\rm F})=e^{-i\frac{2\pi}{3}}$, and
$f_3({\bf k}_{\rm F})=e^{+i\frac{2\pi}{3}}$.
Substituting $f_a({\bf k}_{\rm F})$ into Eq.~(\ref{eq:fa_KF}),
we obtain from Eq.~(\ref{eq:H0_bb}) that 
\begin{align}
 \begin{split}
  & \langle \Psi_{\rm A}^{{\bf k}_{\rm F}+{\bf k}} | 
  {\cal H}_0|\Psi_{\rm B}^{{\bf k}_{\rm F}+{\bf k}} \rangle = 
  \gamma_0 \frac{3a_{\rm cc}}{2} (k_x - ik_y)
  + {\cal O}(k^2), \\
  & \langle \Psi_{\rm B}^{{\bf k}_{\rm F}+{\bf k}} | 
  {\cal H}_0|\Psi_{\rm A}^{{\bf k}_{\rm F}+{\bf k}} \rangle = 
  \gamma_0 \frac{3a_{\rm cc}}{2} (k_x + ik_y)
  + {\cal O}(k^2),
 \end{split}
 \label{eq:H0_K}
\end{align}
where we used $\langle \Psi_{\rm A}^{{\bf k}_{\rm F}} | 
{\cal H}_0|\Psi_{\rm B}^{{\bf k}_{\rm F}} \rangle 
=-\gamma_0 f({\bf k}_{\rm F})=0$.

From Eq.~(\ref{eq:H0_K}), we see that Eq.~(\ref{eq:H_AB}) becomes
\begin{align}
 E({\bf k}_{\rm F}+ {\bf k}) 
 \begin{pmatrix}
  | \Psi_{\rm A}^{{\bf k}_{\rm F}+ {\bf k}} \rangle \cr
  | \Psi_{\rm B}^{{\bf k}_{\rm F}+ {\bf k}} \rangle
 \end{pmatrix}
 = \frac{3 \gamma_0 a_{\rm cc}}{2}
 \begin{pmatrix}
  0 & k_x - ik_y \cr
  k_x + ik_y & 0 
 \end{pmatrix}
 \begin{pmatrix}
  | \Psi_{\rm A}^{{\bf k}_{\rm F}+ {\bf k}} \rangle \cr
  | \Psi_{\rm B}^{{\bf k}_{\rm F}+ {\bf k}} \rangle 
 \end{pmatrix}.
 \label{eq:H_AB_n}
\end{align}
By introducing
the Fermi velocity as 
$v_{\rm F} \equiv \gamma_0 3a_{\rm cc}/2\hbar$
($\sim 10^6$m/s),
the momentum operator $\hat{\bf p}=-i\hbar \nabla$, 
and the Pauli matrix $\bsigma=(\sigma_x,\sigma_y)$,
we obtain the effective-mass Hamiltonian 
near the K point as
\begin{align}
 {\cal H}^{\rm K}_0 = v_{\rm F} \bsigma \cdot \hat{\mathbf{p}}.
 \label{eq:H0K}
\end{align}
${\cal H}^{\rm K}_0$ is a $2\times 2$ matrix 
which operates on the two component wavefunction:
$\psi^{\rm K}({\bf r}) = 
{}^t(\psi^{\rm K}_{\rm A}({\bf r}),\psi^{\rm K}_{\rm B}({\bf r}))$
where $\psi^{\rm K}_{\rm A}({\bf r})$ and $\psi^{\rm K}_{\rm B}({\bf r})$
are the Bloch wavefunction for $\pi$-electrons
of the A and B atom in the unit cell.
By introducing
$\Theta({\bf k})$ which is defined by an angle of 
${\bf k}=(k_x,k_y)$ measured from the $k_x$-axis,
then we get 
$(k_x,k_y) \equiv |{\bf k}| 
(\cos \Theta({\bf k}), \sin \Theta({\bf k}))$,
and
\begin{align}
 {\cal H}_0^{\rm K} =
 \hbar v_{\rm F}|{\bf k}|
 \begin{pmatrix}
  0 & e^{-i\Theta({\bf k})} \cr e^{+i\Theta({\bf k})} & 0
 \end{pmatrix}.
 \label{eq:H0K_k}
\end{align}
The energy eigenvalue of Eq.~(\ref{eq:H0K_k})
is given by $\pm v_{\rm F}|{\bf p}|$ and 
the energy dispersion relation
shows a linear dispersion relation near the K point 
as is known as the Dirac cone.
At the K (or K') point,
the valence and conduction bands touch to each other and thus
we call the degenerated point as the Dirac point.
The eigenstates for $E=+v_{\rm F}|{\bf p}|$ and 
$E=-v_{\rm F}|{\bf p}|$ are given by
\begin{align}
 \psi^{\rm K}_{c,{\bf k}}({\bf r})
 = \frac{e^{i{\bf k}\cdot {\bf r}}}{\sqrt{2S}}
 \begin{pmatrix}
  e^{-i\Theta({\bf k})/2} \cr
  e^{+i\Theta({\bf k})/2}
 \end{pmatrix}, \ \
 \psi^{\rm K}_{v,{\bf k}}({\bf r})
 = \frac{e^{i{\bf k}\cdot {\bf r}}}{\sqrt{2S}}
 \begin{pmatrix}
  e^{-i\Theta({\bf k})/2} \cr
  -e^{+i\Theta({\bf k})/2}
 \end{pmatrix},
 \label{eq:wfK}
\end{align}
which are a conduction band state and a valence band state 
with the wavevector ${\bf k}$.~\footnote{
We note that 
\begin{align}
 \psi^{\rm K}_{c,{\bf k}}({\bf r})
 = \frac{e^{i{\bf k}\cdot {\bf r}}}{\sqrt{2S}}
 \begin{pmatrix}
  1 \cr
  e^{+i\Theta({\bf k})}
 \end{pmatrix}, \ \
 \psi^{\rm K}_{v,{\bf k}}({\bf r})
 = \frac{e^{i{\bf k}\cdot {\bf r}}}{\sqrt{2S}}
 \begin{pmatrix}
  1 \cr
  -e^{+i\Theta({\bf k})}
 \end{pmatrix},
 \label{eq:singl-wf}
\end{align}
are also eigenstates of Eq.~(\ref{eq:H0K})
with an additional phase factor of $e^{\pm \Theta({\bf k})/2}$.
These wavefunctions do not change their values 
under a $2\pi$ rotation in the $k$-space, 
$\Theta({\bf k})\to \Theta({\bf k})+2\pi$.
This behavior is different from Eq.~(\ref{eq:wfK}),
in which the wavefunction of Eq.~(\ref{eq:wfK}) changes the sign 
after the $2\pi$ rotation.
However, the Berry's phase for Eq.~(\ref{eq:singl-wf}) 
which is defined by 
\begin{align}
 \Phi \equiv \int_0^{2\pi}d\Theta
 \left[
 \frac{1}{\sqrt{2}}
 \begin{pmatrix}
  1,e^{-i\Theta({\bf k})}
 \end{pmatrix}
 \right]\left(
 -i
 \frac{\partial}{\partial \Theta}\right)
\left[
 \frac{1}{\sqrt{2}}
 \begin{pmatrix}
  1 \cr
  e^{+i\Theta({\bf k})}
 \end{pmatrix}
\right]=\pi,
\end{align}
gives an extra phase shift of $\pi$.
On the other hand, 
the Berry's phase for Eq.~(\ref{eq:wfK}) vanishes
because
\begin{align}
 \Phi \equiv \int_0^{2\pi}d\Theta
 \left[
 \frac{1}{\sqrt{2}}
 \begin{pmatrix}
  e^{+i\Theta({\bf k})/2},e^{-i\Theta({\bf k})/2}
 \end{pmatrix}
\right]\left(
 -i
 \frac{\partial}{\partial \Theta}\right)
\left[
 \frac{1}{\sqrt{2}}
 \begin{pmatrix}
  e^{-i\Theta({\bf k})/2} \cr
  e^{+i\Theta({\bf k})/2}
 \end{pmatrix}
\right]=0.
\end{align}
Thus, a $2\pi$ rotation of a pseudospin wavefunction
around the K point in the two-dimensional Brillouin zone 
gives minus sign to the wavefunction
in the both cases of Eqs.~(\ref{eq:wfK}) and (\ref{eq:singl-wf}).
It is convenient to use the wavefunction of Eq.~(\ref{eq:wfK}),
since the effect of Berry's phase is included in the wavefunction.
}
In Eq.~(\ref{eq:wfK}),
$S$ denotes the surface area of graphene.
In the case of single wall carbon nanotube (SWNT),
the $k_x$-axis is taken in the circumferential direction 
on the cylindrical surface and the 
$k_y$-axis is defined by the direction 
of a zigzag nanotube axis
(see the coordinate system in Fig.~\ref{fig:graphene}(a)).
We see in Eq.~(\ref{eq:wfK}) that
the energy eigenstate for the valence band,
$\psi^{\rm K}_{v,{\bf k}}({\bf r})$ is given by 
$\sigma_z \psi^{\rm K}_{c,{\bf k}}({\bf r})$.
This is because of 
a particle-hole symmetry of the Hamiltonian:
${\cal H}^{\rm K}_0 \sigma_z 
=- \sigma_z {\cal H}_0^{\rm K}$.~\footnote{
${\cal H}^{\rm K}_0 \sigma_z 
=- \sigma_z {\cal H}_0^{\rm K}$ is obtained directly by
Eqs.~(\ref{eq:H0K}) and $\sigma_z \sigma_i = - \sigma_i \sigma_z$
where $i=x$ or $y$.
Then ${\cal H}^{\rm K}_0 (\sigma_z \psi^{\rm K}_{c,{\bf k}}({\bf r}))
=- \sigma_z {\cal H}_0^{\rm K} \psi^{\rm K}_{c,{\bf k}}({\bf r})
=- \sigma_z v_{\rm F}|{\bf p}| \psi^{\rm K}_{c,{\bf k}}({\bf r})
=(-v_{\rm F}|{\bf p}|) \sigma_z \psi^{\rm K}_{c,{\bf k}}({\bf r})
$.}

Similarly, 
the effective-mass Hamiltonian for the K' point is given by
expanding $f_a({\bf k})$ 
around $-{\bf k}_{\rm F}$ (the K' point) in Eq.~(\ref{eq:H0_bb})
as 
\begin{align}
 {\cal H}^{\rm K'}_0 = v_F \bsigma' \cdot \hat{\mathbf{p}},
 \label{eq:H0_K'}
\end{align}
where $\bsigma' \equiv (-\sigma_x,\sigma_y)$ and 
${\cal H}^{\rm K'}_0$ operates on a two-component 
wavefunction:
$\psi^{\rm K'}({\bf r}) = 
{}^t(\psi^{\rm K'}_{\rm A}({\bf r}),\psi^{\rm K'}_{\rm B}({\bf r}))$.
The energy eigenstates for the conduction and valence energy band
are given, respectively, by
\begin{align}
 \psi^{\rm K'}_{c,{\bf k}'}({\bf r})
 = \frac{e^{i{\bf k}' \cdot {\bf r}}}{\sqrt{2S}}
 \begin{pmatrix}
  e^{+i\Theta({\bf k}')/2} \cr
  -e^{-i\Theta({\bf k}')/2}
 \end{pmatrix}, \ \
 \psi^{\rm K'}_{v,{\bf k}'}({\bf r})
 = \sigma_z \psi^{\rm K'}_{c,{\bf k}'}({\bf r})
 =
\frac{e^{i{\bf k}' \cdot {\bf r}}}{\sqrt{2S}}
 \begin{pmatrix}
  e^{+i\Theta({\bf k}')/2} \cr
  e^{-i\Theta({\bf k}')/2}
 \end{pmatrix}
\end{align}
where $k'_x - i k'_y\equiv |{\bf k}'| e^{-i\Theta({\bf k}')}$.

The linear energy dispersion relations 
near the K and K' points, $E=\pm v_{\rm F}|{\bf p}|$,
are contrasted to 
the non-relativistic energy dispersion relation 
of $|{\bf p}|^2/2m$ where $m$ is the effective-mass of the particle.
The effective-mass for graphene can be understood to be zero
from the definition of relativistic energy 
$E=\pm \sqrt{p^2 c^2 + m^2 c^4}=\pm c|{\bf p}|$ for $m=0$,
where $c$ is substituted to $v_{\rm F}$ ($\sim c/300$).
The wavefunction with two components
is defined by the ``pseudospin''.
The pseudospin up (down) state ${}^t(1,0)$ (${}^t(0,1)$)
corresponds to the wavefunction which has a value only on A (B) atoms.
As we see in Eq.~(\ref{eq:wfK}), if we rotate ${\bf k}$ by $2\pi$
around the K point ${\bf k}=0$, that is,
$\Theta({\bf k})+2\pi$, 
then $\psi_{c,\Theta({\bf k})+2\pi}^{\rm K}({\bf r})
=-\psi_{c,\Theta({\bf k})}^{\rm K}({\bf r})$,
which is the same structure for a real spin under 
a $2\pi$ rotation in the real space.

Here, we give an example to show that the pseudospin 
is relevant to a vanishing matrix element for the lowest 
order backscattering amplitude.~\cite{suzuura02}
When a carrier in the conduction band is denoted by
$| {\bf k}c \rangle$, then
the matrix element of the backscattering process is 
written as $\langle {\bf -k}c | {\cal H}_{\rm imp} | {\bf k}c \rangle$.
When the impurity potential is long-range
compared with the lattice constant, 
the potential is modeled by a diagonal form and 
the matrix element becomes 
\begin{align}
 \langle {\bf -k}c | {\cal H}_{\rm imp} | {\bf k}c \rangle
 = \int \frac{d^2{\bf r}}{2S}
 e^{i2{\bf k}\cdot {\bf r}}
  \begin{pmatrix}
   e^{+i\Theta(-{\bf k})/2} \cr
   e^{-i\Theta(-{\bf k})/2}
 \end{pmatrix}^t
 \begin{pmatrix}
  V({\bf r}) & 0 \cr 0 & V({\bf r})
 \end{pmatrix}
  \begin{pmatrix}
   e^{-i\Theta({\bf k})/2} \cr
   e^{+i\Theta({\bf k})/2}
  \end{pmatrix}
 =0,
\end{align}
where we use Eq.~(\ref{eq:wfK})
and $\Theta({\bf -k})=\Theta({\bf k})\pm \pi$.
This means that 
the interference between the two component pseudospin
makes the back scattering matrix element vanish.
In general, there are many impurities in a sample, 
which gives rise to the Anderson localization in 
the disordered systems.
There is an interesting way to explain the ballistic transport
of carbon nanotube even in the many scattering events.~\cite{ando98}
For a back scattered wave, 
we can find a time-reversal scattering wave whose 
wavefunction has an additional phase shift of $\pi$.
The time reversal pair of scattered waves~\footnote{
Here the time reversal pair of scattering is defined 
for the scatterings within one valley of the K (K') point.
Later, we will use ``time reversal symmetry'' for the pair of ${\bf k}$
around the K point and ${\bf k'}$ around the K' point, when we consider
the intervalley scattering.
} cancel with each
other, which leads to the absence of the backward
scattering.~\cite{ando98}
This is contrasted with the standard concept that impurity
scattering gives rise to the Anderson localization for
low-dimensional systems.

\section{Gauge Field for a Deformed Graphene}\label{sec:weyl_deform}

The dynamics of the conducting electrons in graphene materials
are different from those of ideal flat graphene,
because in the former case, 
there are shape fluctuations 
(cylindrical shape, phonon vibration, etc.) 
that result in the modification of the overlap matrix elements.
of nearest-neighbor $\pi$-orbitals and 
of the on-site potential energy.
We refer the modification of the nearest-neighbor hopping integral 
as the off-site interaction and
a shift of the on-site potential energy as
the on-site interaction.
The modification of the effective-mass equations 
by the off-site (on-site) interaction is discussed
in Sec.~\ref{sub2sec:off} (Sec.~\ref{sub2sec:on}).
We discuss time-reversal symmetry property of the 
effective-mass equations 
including the off-site and on-site interactions
in Sec.~\ref{sub2sec:sym}.

\subsection{Off-site interaction}\label{sub2sec:off}

First we consider the perturbation from 
the off-site interaction in which 
only off-diagonal matrix element has a non-zero value.
A lattice deformation induces 
a local modification of the nearest-neighbor hopping integral
as $-\gamma_0 \to -\gamma_0+\delta \gamma_0^a({\bf r}_i)$ 
($a=1,2,3$) (see Fig.~\ref{fig:graphene}).
We define the perturbation ${\cal H}_1$ as 
\begin{align}
 {\cal H}_1 \equiv 
 \sum_{i \in {\rm A}} \sum_{a=1,2,3} 
 \delta \gamma^a_0(\mathbf{r}_i) 
 \left(
 (c_{i+a}^{\rm B})^\dagger c_i^{\rm A} + 
 (c_i^{\rm A})^\dagger c_{i+a}^{\rm B}
 \right).
 \label{eq:H1}
\end{align}
The off-site matrix element of ${\cal H}_1$ 
with respect to the Bloch wave functions
in Eq.~(\ref{eq:Bloch}) with ${\bf k}$
and ${\bf k}+\delta {\bf k}$
is given by
\begin{align}
 \begin{split}
  & \langle \Psi_{\rm A}^{{\bf k}+\delta {\bf k}} |{\cal H}_1|
  \Psi_{\rm B}^{{\bf k}} \rangle
  = \frac{1}{N_u} \sum_{i \in {\rm A}} \sum_{a=1,2,3} 
  \delta \gamma^a_0({\bf r}_i) f_a({\bf k}) e^{-i\delta {\bf k}
  \cdot {\bf r}_i}, \\
  & \langle \Psi_{\rm B}^{{\bf k}+\delta {\bf k}} |{\cal H}_1|
  \Psi_{\rm A}^{{\bf k}} \rangle 
  = \frac{1}{N_u} \sum_{i \in {\rm A}} \sum_{a=1,2,3} 
  \delta \gamma^a_0({\bf r}_i) f_a({\bf k})^* 
  e^{-i\delta {\bf k}\cdot ({\bf r}_i+{\bf R}_a)}.
 \end{split}
 \label{eq:H1_bb}
\end{align}
Here we consider the two possible cases
for $\delta {\bf k}$.
When $\delta {\bf k}$ is sufficiently small
compared with the reciprocal lattice vector,
${\bf k}$ near the K (or K') point
is scattered to the ${\bf k'}={\bf k}+\delta {\bf k}$
within the region near the K (or K') point,
which we call the intravalley scattering.
On the other hand,
if $\delta {\bf k}$ is comparable
to the distance between the K and K' points,
we expect scattering from K (K') point to K' (K) point, 
which we call the intervalley scattering.

\subsubsection{Intravalley Scattering}

In the case of intravalley scattering,
when ${\bf k}$ is measured from ${\bf k}_{\rm F}$,
we get
\begin{align}
 \begin{split}
  & \langle \Psi_{\rm A}^{{\bf k}_{\rm F}+{\bf k}+\delta {\bf k}} |
  {\cal H}_1| \Psi_{\rm B}^{{\bf k}_{\rm F}+{\bf k}} \rangle 
  = \frac{1}{N_u} \sum_{i \in {\rm A}} \sum_{a=1,2,3} 
  \delta \gamma_a({\bf r}_i) f_a({\bf k}_{\rm F}) 
  e^{-i\delta {\bf k} \cdot {\bf r}_i}+
  {\cal O}(\delta k \delta \gamma_a), \\
  & \langle \Psi_{\rm B}^{{\bf k}_{\rm F}+{\bf k}+\delta {\bf k}} |
  {\cal H}_1| \Psi_{\rm A}^{{\bf k}_{\rm F}+{\bf k}} \rangle
  = \frac{1}{N_u} \sum_{i \in {\rm A}} \sum_{a=1,2,3} 
  \delta \gamma_a({\bf r}_i) f_a({\bf k}_{\rm F})^*
  e^{-i\delta {\bf k} \cdot {\bf r}_i}+
  {\cal O}(\delta k \delta \gamma_a),
 \end{split}
 \label{eq:g-basic}
\end{align}
The correction indicated by ${\cal O}(\delta k \delta \gamma_a)$ 
in Eq.~(\ref{eq:g-basic})
is negligible when 
$|\delta {\bf k}| \ll |{\bf k}_{\rm F}|$.
Substituting $f_1({\bf k}_{\rm F}) = 1$, 
$f_2({\bf k}_{\rm F})=e^{-i\frac{2\pi}{3}}$ and 
$f_3({\bf k}_{\rm F})=e^{+i\frac{2\pi}{3}}$ 
into Eq.(\ref{eq:g-basic}),
we get 
\begin{align}
 \begin{split}
  & \langle \Psi_{\rm A}^{{\bf k}_{\rm F}+{\bf k} + \delta {\bf k}} 
  |{\cal H}_1| \Psi_{\rm B}^{{\bf k}_{\rm F}+{\bf k}} \rangle
  = \frac{v_{\rm F}}{N_{u}} \sum_{i \in {\rm A}}
  \left\{
  A_x^{\rm q}({\bf r}_i) -i A_y^{\rm q}({\bf r}_i) \right\}
  e^{-i\delta {\bf k} \cdot {\bf r}_i}, \\
  & \langle \Psi_{\rm B}^{{\bf k}_{\rm F}+{\bf k} + \delta {\bf k}} 
  |{\cal H}_1| \Psi_{\rm A}^{{\bf k}_{\rm F}+{\bf k}} \rangle
  = \frac{v_{\rm F}}{N_{u}} \sum_{i \in {\rm A}}
  \left\{
  A_x^{\rm q}({\bf r}_i) +i A_y^{\rm q}({\bf r}_i) \right\}
  e^{-i\delta {\bf k} \cdot {\bf r}_i},
 \end{split}
 \label{eq:off-ele}
\end{align}
where
${\bf A}^{\rm q}({\bf r})=(A_x^{\rm q}({\bf r}),A_y^{\rm q}({\bf r}))$
is defined by $\delta \gamma^a_0({\bf r})$ ($a=1,2,3$) as
\begin{align}
 \begin{split}
  & v_{\rm F} A_x^{\rm q}({\bf r}) = \delta \gamma^1_0({\bf r})
  - \frac{1}{2} \left( \delta \gamma^2_0({\bf r}) +
  \delta \gamma^3_0({\bf r}) \right), \\
  & v_{\rm F} A_y^{\rm q}({\bf r}) = \frac{\sqrt{3}}{2} 
  \left( \delta \gamma^2_0({\bf r}) -
  \delta \gamma^3_0({\bf r}) \right).
 \end{split}
 \label{eq:A}
\end{align}
Since the diagonal term vanishes, that is,
$\langle \Psi_s^{\bf k}|{\cal H}_1| \Psi_s^{\bf k'}\rangle = 0$
($s={\rm A,B}$),
Eq.~(\ref{eq:off-ele}) shows that ${\cal H}_1$ 
is expressed by 
$v_{\rm F} \bsigma \cdot {\bf A}^{\rm q}({\bf r})$
in the effective-mass Hamiltonian.
Thus, the total Hamiltonian of a deformed graphene 
is expressed by
\begin{align}
 {\cal H}^{\rm K}_0 + {\cal H}^{\rm K}_1
 = v_{\rm F} 
 \bsigma \cdot (\hat{\bf p}+{\bf A}^{\rm q}({\bf r})).
 \label{eq:HK}
\end{align}
Similarly, 
by expanding $f_a({\bf k})$ in Eq.~(\ref{eq:H1_bb}) 
around $-{\bf k}_{\rm F}$ of the K' point, 
we find ${\cal H}_1$ appears as 
$-v_{\rm F} \bsigma' \cdot {\bf A}^{\rm q}({\bf r})$.
Thus, 
we obtain the effective-mass Hamiltonian
for the K' point as
\begin{align}
 {\cal H}^{\rm K'}_0 + {\cal H}^{\rm K'}_1
 = v_{\rm F} 
 \bsigma' \cdot (\hat{\mathbf{p}}-\mathbf{A}^{\rm q}({\bf r})).
 \label{eq:HK'}
\end{align}
The corresponding Schr\"odinger equation for
Eqs~(\ref{eq:HK}) and (\ref{eq:HK'}) are given in 
Eq.~(\ref{eq:weyl_graphene}).~\cite{sasaki05}


Thus, the off-site interaction
can be included in the effective-mass equations 
as a gauge field, ${\bf A}^{\rm q}({\bf r})$.
We call ${\bf A}^{\rm q}({\bf r})$ 
the {\bf deformation-induced gauge field} and distinguish it
from the electromagnetic gauge field ${\bf A}({\bf r})$.~\cite{sasaki05}
${\cal H}_0 + {\cal H}_1$ 
does not break time-reversal symmetry
even tough the ${\bf A}^{\rm q}({\bf r})$ appears as a gauge field,
because the sign in front of ${\bf A}^{\rm q}({\bf r})$ 
is opposite to each other for the K and K' points.
This contrasts with the fact that 
${\bf A}({\bf r})$ (magnetic field)
violates time-reversal symmetry
because the sign in front of ${\bf A}({\bf r})$ 
is the same for the K and K' points
as ${\hat {\bf p}} \to {\hat {\bf p}}-e{\bf A}({\bf r})$.
The time-reversal symmetry 
will be discussed in detail in Sec.~\ref{sub2sec:sym}.
The rotation of ${\bf A}^{\rm q}({\bf r})$ or 
the deformation-induced {\it magnetic} field is defined by 
${\bf B}^{\rm q}({\bf r})= \nabla \times {\bf A}({\bf r})$ as
\begin{align}
 B^{\rm q}_z({\bf r}) 
 = \frac{\partial A_y^{\rm q}({\bf r})}{\partial x}
 -\frac{\partial A_x^{\rm q}({\bf r})}{\partial y}.
 \label{eq:mag-field}
\end{align}
The direction of $B^{\rm q}_z(\mathbf{r})$
is perpendicular to the graphene plane
${\bf B}^{\rm q}(\mathbf{r})= (0,0,B^{\rm q}_z(\mathbf{r}))$.
The deformation-induced {\it magnetic} field 
changes the energy band structure, which
will be shown in \S~\ref{sec:edge}.
The $B^{\rm q}_z({\bf r})$ couples with the pseudospin
through $\sigma_z$ term.
This is shown by taking 
the square of the effective-mass equations,
as we see in Eqs.~(\ref{eq:square_weyl}) and (\ref{eq:square_weyl_2}),
\begin{align}
 \begin{split}
  & E^2 \psi^{\rm K}({\bf r}) =
  v_{\rm F}^2 \left\{ (\hat{\mathbf{p}}+\mathbf{A}^{\rm
  q}(\mathbf{r}))^2 + \hbar 
  B_z^{\rm q}(\mathbf{r}) \sigma_z \right\} \psi^{\rm K}({\bf r}), \\
  & E^2 \psi^{\rm K'}({\bf r}) =
  v_{\rm F}^2 \left\{ (\hat{\mathbf{p}}-\mathbf{A}^{\rm
  q}(\mathbf{r}))^2 + \hbar 
  B^{\rm q}_z(\mathbf{r}) \sigma_z \right\} \psi^{\rm K'}({\bf r}), 
 \end{split}
 \label{eq:H^2}
\end{align}
where $B^{\rm q}_z({\bf r})\sigma_z$ terms modify 
the direction of the pseudospin to the direction 
opposite to $B^{\rm q}_z({\bf r})$ 
in order to decrease the energy
for the both K and K' points.
Here let us stress
a notable difference between $B^{\rm q}_z({\bf r})$
and the usual magnetic field, $B_z({\bf r})$.
For $B_z^{\rm q}({\bf r})$,
a uniform magnetic field along the z-axis over a sample
is possible to realize the uniform spin polarization (the Zeeman term).
However, since $B^{\rm q}_z({\bf r})$ is generated by the modification
of $\gamma_0$,
the field $B^{\rm q}_z({\bf r})$
appears only near the defect region without periodic symmetry
such as the zigzag edge~\cite{sasaki06jpsj}
which gives local polarization of pseudospin.
In general, when a graphene sample contains 
an equal number of A and B atoms to each other, we have 
\begin{align}
 \int_S B^{\rm q}_z({\bf r}) d^2{\bf r} = 0,
\end{align}
where $S$ denotes an integral for a whole sample.

\subsubsection{Intervalley Scattering}

Next, we consider
the matrix element of ${\cal H}_1$ 
for the intervalley scattering.
Let us define the wave vector of a Bloch state 
near the {\rm K} point; ${\bf k}_{\rm F}+{\bf k}$
and that of another Bloch state near the {\rm K'} point;
$-{\bf k}_{\rm F}+{\bf k'}$
($|{\bf k}|,|{\bf k'}| \ll |{\bf k}_{\rm F}|$), 
then the matrix element from the K' to K points
is written by using Eq.~(\ref{eq:H1_bb}) as
\begin{align}
 \langle \Psi_{\rm A}^{{\bf k}_{\rm F}+{\bf k}} |
 {\cal H}_1|
 \Psi_{\rm B}^{-{\bf k}_{\rm F} + {\bf k}'} \rangle
 &= \frac{1}{N_u} \sum_{i \in {\rm A}} \sum_{a=1,2,3} 
 \delta \gamma^a_0(\mathbf{r}_i) f_a({\bf k}_{\rm F})^*
 e^{-i2 {\bf k}_{\rm F} \cdot {\bf r}_i} 
 e^{i({\bf k'}-{\bf k})\cdot {\bf r}_i } 
 e^{i{\bf k'}\cdot {\bf R}_a}
 \nn \\
 & \simeq \frac{1}{N_u} \sum_{i \in {\rm A}} \sum_{a=1,2,3} 
 \delta \gamma^a_0(\mathbf{r}_i) f_a({\bf k}_{\rm F})^*
 e^{-i2 {\bf k}_{\rm F} \cdot {\bf r}_i}
 e^{i({\bf k'}-{\bf k})\cdot {\bf r}_i } \nn \\
 & =\frac{v_{\rm F}}{N_u} \sum_{i \in {\rm A}} 
 \left\{ A_x^{\rm q}({\bf r}_i) +i A_y^{\rm q}( {\bf r}_i) \right\} 
 e^{-i2 {\bf k}_{\rm F} \cdot {\bf r}_i}
 e^{i({\bf k'}-{\bf k})\cdot {\bf r}_i },
 \label{eq:KA_K'B}
\end{align}
and
\begin{align}
 \langle \Psi_{\rm B}^{{\bf k}_{\rm F} + {\bf k}} |
 {\cal H}_1|
 \Psi_{\rm A}^{ -{\bf k}_{\rm F} + {\bf k}'} \rangle 
 &= \frac{1}{N_u} \sum_{i \in {\rm A}} \sum_{a=1,2,3} 
 \delta \gamma^a_0(\mathbf{r}_i) f_a({\bf k}_{\rm F})^*
 e^{-i2 {\bf k}_{\rm F} \cdot {\bf r}_i} 
 e^{i({\bf k'}-{\bf k})\cdot {\bf r}_i } 
 e^{-i{\bf k}\cdot {\bf R}_a}
 \nn \\
 & \simeq \frac{1}{N_u} \sum_{i \in {\rm A}} \sum_{a=1,2,3} 
 \delta \gamma^a_0(\mathbf{r}_i) f_a({\bf k}_{\rm F})^* 
 e^{-i2 {\bf k}_{\rm F} \cdot {\bf r}_i}
 e^{i({\bf k'}-{\bf k})\cdot {\bf r}_i } \nn \\
 & = \frac{v_{\rm F}}{N_u} \sum_{i \in {\rm A}} 
 \left\{ A_x^{\rm q}({\bf r}_i) +i A_y^{\rm q}({\bf r}_i) \right\}
 e^{-i2 {\bf k}_{\rm F} \cdot {\bf r}_i} 
 e^{i({\bf k'}-{\bf k})\cdot {\bf r}_i }.
 \label{eq:KB_K'A}
\end{align}
In Eqs.~(\ref{eq:KA_K'B}) and (\ref{eq:KB_K'A}),
from the first line of the right-hand side to the second line,
we approximate 
$e^{i{\bf k'}\cdot {\bf R}_a}\approx 1$ and 
$e^{-i{\bf k}\cdot {\bf R}_a}\approx 1$
since we consider the case of 
$|{\bf k}'| \ll |{\bf k}_{\rm F}|$ and 
$|{\bf k}| \ll |{\bf k}_{\rm F}|$.
Equations~(\ref{eq:KA_K'B}) and (\ref{eq:KB_K'A})
show that the deformation-induced gauge field 
${\bf A}^{\rm q}({\bf r})$ (Eq.~(\ref{eq:A}))
can be applied to the matrix element 
for intervalley scattering, too.
It is noted that 
Eqs.~(\ref{eq:KA_K'B}) and (\ref{eq:KB_K'A}) 
are identical to each other up to the leading term.
The matrix elements from the K point to the K' point
are given by replacing ${\bf k}_{\rm F}$ by $-{\bf k}_{\rm F}$
in Eqs.~(\ref{eq:KA_K'B}) and (\ref{eq:KB_K'A})
as 
\begin{align}
 \begin{split}
  & \langle \Psi_{\rm A}^{-{\bf k}_{\rm F}+{\bf k}'} |{\cal H}_1|
  \Psi_{\rm B}^{{\bf k}_{\rm F} + {\bf k}} \rangle
  = \frac{v_{\rm F}}{N_u} \sum_{i \in {\rm A}} 
  \left\{ A_x^{\rm q}({\bf r}_i)-i A_y^{\rm q}({\bf r}_i) \right\}
  e^{+i2 {\bf k}_{\rm F} \cdot {\bf r}_i} 
  e^{-i({\bf k'}-{\bf k})\cdot {\bf r}_i }, \\
  & \langle \Psi_{\rm B}^{-{\bf k}_{\rm F}+{\bf k}'} |{\cal H}_1|
  \Psi_{\rm A}^{{\bf k}_{\rm F}+{\bf k}} \rangle
  = \frac{v_{\rm F}}{N_u} \sum_{i \in {\rm A}} 
  \left\{ A_x^{\rm q}({\bf r}_i)-i A_y^{\rm q}({\bf r}_i) \right\}
  e^{+i2 {\bf k}_{\rm F} \cdot {\bf r}_i} 
  e^{-i({\bf k'}-{\bf k})\cdot {\bf r}_i }.
 \end{split}
\end{align}

In the effective-mass model,
the intervalley scattering appears as the off-diagonal terms
in a $4\times 4$ matrix:
\begin{align}
 E 
 \begin{pmatrix}
  \psi^{\rm K}({\bf r}) \cr \psi^{\rm K'}({\bf r})
 \end{pmatrix}= 
 v_{\rm F}
 \begin{pmatrix}
  \bsigma \cdot (\hat{\bf p}+{\bf A}^{\rm q}({\bf r})) 
  & \phi^{\rm q}({\bf r}) \sigma_x \cr
  \phi^{\rm q}({\bf r})^* \sigma_x
  & \bsigma' \cdot (\hat{\bf p}-{\bf A}^{\rm q}({\bf r}))
 \end{pmatrix}
 \begin{pmatrix}
  \psi^{\rm K}({\bf r}) \cr \psi_{\rm K'}({\bf r})
 \end{pmatrix},
 \label{eq:Weyl_KK'}
\end{align}
where we define a deformation-induced 
complex scalar field $\phi({\bf r})$
\begin{align}
 \phi^{\rm q}({\bf r})\equiv 
 (A_x^{\rm q}({\bf r})+i A_y^{\rm q}({\bf r}))e^{-i2{\bf k}_{\rm F}\cdot{\bf r}}.
 \label{eq:phi}
\end{align}
When ${\bf A}^{\rm q}({\bf r})$ is a smooth function of ${\bf r}$, 
the effect of intervalley scattering by the complex scalar field 
$\phi^{\rm q}({\bf r})$ can be neglected because 
$\phi^{\rm q}({\bf r})$ becomes a rapidly oscillating function of ${\bf r}$
due to the factor $e^{-i2{\bf k}_{\rm F}\cdot{\bf r}}$,
and the value of matrix element between the K and K' points 
becomes negligible, too.
On the other hand, 
when ${\bf A}^{\rm q}({\bf r})$ 
is a rapidly oscillating function of ${\bf r}$,
$\phi^{\rm q}({\bf r})$ becomes a smooth function of ${\bf r}$
and the matrix element between the K and K' points is not negligible.
For example, phonon modes at the K point contribute to the
intervalley scattering. 
Since the vibrational amplitude of the K point phonon modes
are proportional to $e^{-i{\bf k}_{\rm F}\cdot {\bf r}_i}$,
${\bf A}^{\rm q}({\bf r}_i)$ has the same periodicity of 
$e^{+2i{\bf k}_{\rm F}\cdot {\bf r}_i}$ where
we use the fact $e^{+3i{\bf k}_{\rm F}\cdot {\bf r}_i}=1$.
Then, $\phi^{\rm q}({\bf r}_i)$ would behave as a constant in 
the real space.

\subsection{On-site interaction}\label{sub2sec:on}

Now we consider the on-site interaction by
a defect of the crystal.
A lattice deformation gives rise to 
not only a change of the transfer integral 
between {\rm A} and {\rm B} atoms
but also a change of the potential at the A (B) atom
$\phi_A$ ($\phi_B$) which we call the off-site and 
on-site deformation potential, respectively.
We denote the on-site deformation potential 
by a $2 \times 2$ matrix for 
$\psi={}^t(\psi^{\rm K},\psi^{\rm K'})$
\begin{align}
 {\cal H}_{\rm on} 
 =
 \begin{pmatrix}
  \phi_{\rm A}({\bf r}_i) & 0 \cr
  0 & \phi_{\rm B}({\bf r}_i+{\bf R}_1)
 \end{pmatrix}.
 \label{eq:phi_AB}
\end{align}
Let the displacement vector of {\rm A}-atom at ${\bf r}_i$
is ${\bf u}_{\rm A}({\bf r}_i)$
and that of {\rm B}-atom at ${\bf r}_j$
is ${\bf u}_{\rm B}({\bf r}_j)$, then
the deformation potential of {\rm A}-atom at ${\bf r}_i$,
$\phi_{\rm A}({\bf r}_i)$, 
is induced by the relative 
displacements of three nearest neighbor B-atoms from the {\rm A}-atom 
(${\bf u}_{\rm B}({\bf r}_i + {\bf R}_a) - {\bf u}_{\rm A}({\bf r}_i)$)
as 
\begin{align}
 \phi_{\rm A}({\bf r}_i) = \frac{g_{\rm on}}{\ell a_{\rm cc}}
 \sum_{a=1,2,3}
 {\bf R}_a \cdot 
 \left(
 {\bf u}_{\rm B}({\bf r}_i + {\bf R}_a) - {\bf u}_{\rm A}({\bf r}_i)
 \right),
 \label{eq:phi_A}
\end{align}
where $g_{\rm on}$ denotes gradient of the atomic potential
at ${\bf r}_i$, and $\ell$ denotes $3a_{\rm cc}/2$.
Here we assume that
$|{\bf u}_{\rm B}({\bf r}_i + {\bf R}_a) - {\bf u}_{\rm A}({\bf
r}_i)|\ll a_{\rm cc}$ and that 
$\phi_{\rm A}({\bf r}_i)$ depends linearly
on the relative displacement vector.
\begin{figure}[htbp]
 \begin{center}
  \includegraphics[scale=0.7]{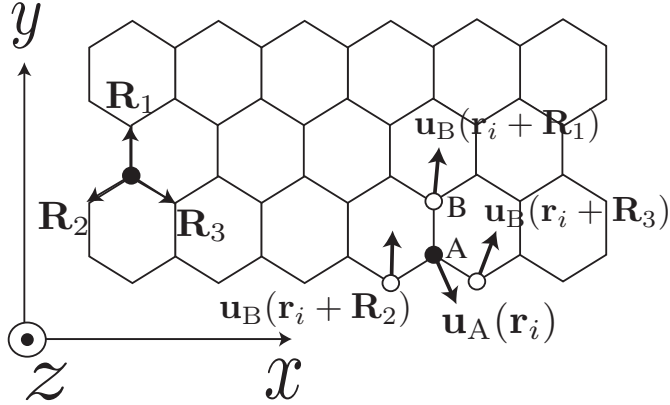}
 \end{center}
 \caption{
 Displacements of B-atoms at ${\bf r}_i + {\bf R}_a$ ($a=1,2,3$),
 ${\bf u}_{\rm B}({\bf r}_i + {\bf R}_a)$,
 give rise to a deformation potential at A-atom of ${\bf r}_i$.
 }
 \label{fig:onsite}
\end{figure}

By expanding ${\bf u}_{\rm B}({\bf r}_i + {\bf R}_2)$ as 
${\bf u}_{\rm B}({\bf r}_i + {\bf R}_2)=
{\bf u}_{\rm B}({\bf r}_i + {\bf R}_1)+(({\bf R}_2-{\bf R}_1)\cdot
\nabla){\bf u}_{\rm B}({\bf r}_i + {\bf R}_1)+\cdots$
and ${\bf u}_{\rm B}({\bf r}_i + {\bf R}_3)$ as 
${\bf u}_{\rm B}({\bf r}_i + {\bf R}_3)=
{\bf u}_{\rm B}({\bf r}_i + {\bf R}_1)+(({\bf R}_3-{\bf R}_1)\cdot
\nabla){\bf u}_{\rm B}({\bf r}_i + {\bf R}_1)+\cdots$,
we see that Eq.~(\ref{eq:phi_A}) can be approximated by
\begin{align}
 \phi_{\rm A}({\bf r}_i)=
 g_{\rm on} \nabla \cdot {\bf u}_{\rm B}({\bf r}_i + {\bf R}_1) +\cdots,
 \label{eq:phi_A_smo}
\end{align}
where we have used $\sum_{a=1,2,3}{\bf R}_a=0$.
It is noted that a general expression for the deformation potential,
Eq.~(\ref{eq:phi_A_smo}), is valid 
in the case that ${\bf u}_{\rm B}({\bf r})$
is a smooth function of ${\bf r}$.
When this is not the case,
we have to use Eq.~(\ref{eq:phi_A}).
In the continuous limit,
we may use ${\bf r}$ to represent 
the positions of both {\rm A} and {\rm B} atoms
in the unit cell, 
then we have $\phi_{\rm A}({\bf r})= g_{\rm on} \nabla \cdot {\bf u}_{\rm
B}({\bf r}) +\cdots$. 
Similarly, 
the deformation potential of {\rm B}-site 
of ${\bf r}_i + {\bf R}_1$ is given by
\begin{align}
 \phi_{\rm B}({\bf r}_i+{\bf R}_1) = 
 \frac{g_{\rm on}}{\ell a_{\rm cc}}
 \sum_{a=1,2,3}
 -{\bf R}_a \cdot 
 \left(
 {\bf u}_{\rm A}({\bf r}_i+{\bf R}_1- {\bf R}_a)-
 {\bf u}_{\rm B}({\bf r}_i+{\bf R}_1) 
 \right).
 \label{eq:phi_B}
\end{align}
By using
${\bf u}_{\rm A}({\bf r}_i + {\bf R}_1-{\bf R}_2)=
{\bf u}_{\rm A}({\bf r}_i)+(({\bf R}_1-{\bf R}_2)\cdot
\nabla){\bf u}_{\rm A}({\bf r}_i)+\cdots$
and 
${\bf u}_{\rm A}({\bf r}_i + {\bf R}_1-{\bf R}_3)=
{\bf u}_{\rm A}({\bf r}_i)+(({\bf R}_1-{\bf R}_3)\cdot
\nabla){\bf u}_{\rm A}({\bf r}_i)+\cdots$,
we see that Eq.~(\ref{eq:phi_B}) can be approximated by
\begin{align}
 \phi_{\rm B}({\bf r}_i + {\bf R}_1)=
 g_{\rm on} \nabla \cdot {\bf u}_{\rm A}({\bf r}_i) +\cdots. 
 \label{eq:phi_B_smo}
\end{align}
Thus, for the intravalley scattering, 
we may rewrite Eq.~(\ref{eq:phi_AB}) using
Eqs.~(\ref{eq:phi_A_smo}) and (\ref{eq:phi_B_smo}) as 
\begin{align}
 {\cal H}_{\rm on} = g_{\rm on}
 \begin{pmatrix}
  \nabla \cdot {\bf u}_{\rm B}({\bf r}) & 0 \cr
  0 & \nabla \cdot {\bf u}_{\rm A}({\bf r})
 \end{pmatrix}
 +\cdots.
 \label{eq:Hon_g}
\end{align}
According to the result of density-functional theory
by Porezag {\it et al.},~\cite{porezag95,sasaki07local} 
we will use the parameter for $g_{\rm on}$ ($=$17eV).
For the later discussion of el-ph interaction of 
acoustic
\begin{align}
 {\bf s}({\bf r})
 \equiv \frac{{\bf u}_{\rm A}({\bf r})+{\bf u}_{\rm B}({\bf r})}{2},
\end{align}
and optical 
\begin{align}
 {\bf u}({\bf r})
 \equiv {\bf u}_{\rm B}({\bf r})-{\bf u}_{\rm A}({\bf r}),
 \label{eq:u_opt}
\end{align}
phonon modes,
we rewrite Eq.~(\ref{eq:Hon_g}) 
using the Pauli matrices as
\begin{align}
 {\cal H}_{\rm on} 
 = \frac{g_{\rm on}}{2} 
 I
 \nabla \cdot 
 ({\bf u}_{\rm A}({\bf r})+{\bf u}_{\rm B}({\bf r}))
 + \frac{g_{\rm on}}{2} \sigma_z
 \nabla \cdot 
 ({\bf u}_{\rm B}({\bf r})-{\bf u}_{\rm A}({\bf r})).
 \label{eq:onsite_psin}
\end{align}
Comparing with the expressions for 
${\bf A}^{\rm q}({\bf r})$ and $\phi^{\rm q}({\bf r})$
in Eqs.~(\ref{eq:A}) and (\ref{eq:phi}), and for 
Eq.~(\ref{eq:onsite_psin}),
the effective-mass Hamiltonian
of the defect of the crystal is given by
\begin{align}
 {\cal H}= v_{\rm F}
 \begin{pmatrix}
  {\cal H}^{\rm K}
  & v_{\rm F} \phi^{\rm q}({\bf r}) \sigma_x \cr
  v_{\rm F} \phi^{\rm q}({\bf r})^* \sigma_x
  & {\cal H}^{\rm K'}
 \end{pmatrix},
\end{align}
where we define
\begin{align}
 \begin{split}
  & {\cal H}^{\rm K} = v_{\rm F}\bsigma \cdot (\hat{\bf p}+{\bf A}^{\rm q}({\bf r})) 
  + g_{\rm on} I \nabla \cdot {\bf s}({\bf r})
 + \frac{g_{\rm on}}{2} \sigma_z \nabla \cdot 
 {\bf u}({\bf r}), \\
  & {\cal H}^{\rm K'} =v_{\rm F}\bsigma' \cdot (\hat{\bf p}-{\bf A}^{\rm q}({\bf r}))
  + g_{\rm on} I \nabla \cdot {\bf s}({\bf r})
 + \frac{g_{\rm on}}{2} \sigma_z \nabla \cdot 
 {\bf u}({\bf r}).
 \end{split}
\end{align}

\subsection{Time-reversal and pseudospin symmetries}\label{sub2sec:sym}

Here we show that 
${\bf A}^{\rm q}({\bf r})$ keeps time-reversal symmetry 
of graphene system.
The time-reversal operation, ${\cal T}$, is defined by 
exchanging the K point and the K' point, and 
taking a complex conjugation of the four component
wavefunction as
\begin{align}
 {\cal T}
 \begin{pmatrix}
  \psi^{\rm K}({\bf r}) \cr \psi^{\rm K'}({\bf r})
 \end{pmatrix}
 =
 \begin{pmatrix}
  \psi^{\rm K'}({\bf r})^* \cr \psi^{\rm K}({\bf r})^*
 \end{pmatrix}, \ {\rm where} \
 {\cal T}=
 \begin{pmatrix}
  0 & I \cr
  I & 0 
 \end{pmatrix}
 K
 \label{eq:trs}
\end{align}
and $K$ is the complex conjugate operator.
In order to check that Eq.~(\ref{eq:trs}) 
is time-reversal operation,
it is useful to introduce the vector potential 
for electro-magnetism, ${\bf A}({\bf r})$,
into the effective-mass Hamiltonian as
\begin{align}
 {\cal H}
 = v_{\rm F}
 \begin{pmatrix}
  \bsigma \cdot (\hat{\bf p}+{\bf A}^{\rm q}({\bf r})-e{\bf A}({\bf r})) 
  & \phi^{\rm q}({\bf r}) \sigma_x \cr
  \phi^{\rm q}({\bf r})^* \sigma_x
  & \bsigma' \cdot (\hat{\bf p}-{\bf A}^{\rm q}({\bf r})-e{\bf A}({\bf r}))
 \end{pmatrix}.
 \label{eq:Weyl_KK'_Aem}
\end{align}
Then the electromagnetic current operator, 
${\bf J}({\bf r})$, is given by
\begin{align}
 {\bf J}({\bf r})
 = \frac{\partial {\cal H}}{\partial {\bf A}({\bf r})}
 = -e v_{\rm F} 
 \begin{pmatrix}
  \bsigma & 0 \cr 0 & \bsigma'
 \end{pmatrix}.
\end{align}
Using ${\cal T}$ in Eq.~(\ref{eq:trs}), 
we can show that
\begin{align}
 {\cal T}^{-1}{\bf J}({\bf r}){\cal T}=
 -{\bf J}({\bf r}), 
 \label{eq:J_em_T}
\end{align}
where we have used $K^{-1} \bsigma K = - \bsigma'$
and $K^{-1} \bsigma' K = - \bsigma$.
The negative sign on the right-hand side of Eq.~(\ref{eq:J_em_T})
shows that Eq.~(\ref{eq:trs}) is time-reversal operation.

When we apply ${\cal T}$ to the wavefunction of
Eq.~(\ref{eq:Weyl_KK'}), we get
\begin{align}
 E 
 \begin{pmatrix}
  \psi^{\rm K'}({\bf r})^* \cr \psi^{\rm K}({\bf r})^*
 \end{pmatrix}
 =
 v_{\rm F}
 \begin{pmatrix}
  \bsigma \cdot (\hat{\bf p}+{\bf A}^{\rm q}({\bf r})) 
  & \phi^{\rm q}({\bf r}) \sigma_x \cr
  \phi^{\rm q}({\bf r})^* \sigma_x
  & \bsigma' \cdot (\hat{\bf p}-{\bf A}^{\rm q}({\bf r}))
 \end{pmatrix}
 \begin{pmatrix}
  \psi^{\rm K'}({\bf r})^* \cr \psi^{\rm K}({\bf r})^*
 \end{pmatrix}.
 \label{eq:Weyl_T}
\end{align}
By taking the complex conjugation (c.c.)
of the first row of Eq.~(\ref{eq:Weyl_T}),
we get
\begin{align}
 & E \psi^{\rm K'}({\bf r})^* =
 v_{\rm F}\bsigma \cdot 
 (\hat{\bf p}+{\bf A}^{\rm q}({\bf r}))\psi^{\rm K'}({\bf r})^*
 + v_{\rm F}\phi^{\rm q}({\bf r}) \sigma_x \psi^{\rm K}({\bf r})^*
 \nn \\
 & \stackrel{\rm c.c.}{\to}
 E \psi^{\rm K'}({\bf r}) =
 v_{\rm F}\bsigma' \cdot (\hat{\bf p}-{\bf A}^{\rm q}({\bf r}))
 \psi^{\rm K'}({\bf r})
 + v_{\rm F}\phi^{\rm q}({\bf r})^* \sigma_x \psi^{\rm K}({\bf r})
 \label{eq:psi_K_T}
\end{align}
where we have used 
$\bsigma^*=-\bsigma'$ and $\hat{\bf p}^* = -\hat{\bf p}$.
Similarly, by taking the complex conjugation 
of the second row of Eq.~(\ref{eq:Weyl_T}),
we get
\begin{align}
 &
 E \psi^{\rm K}({\bf r})^* =
 v_{\rm F}\phi^{\rm q}({\bf r})^* \sigma_x \psi^{\rm K'}({\bf r})^*
 +v_{\rm F}\bsigma' \cdot 
 (\hat{\bf p}-{\bf A}^{\rm q}({\bf r}))\psi^{\rm K}({\bf r})^*
 \nn \\
 & \stackrel{\rm c.c.}{\to}
 E \psi^{\rm K}({\bf r}) =
 v_{\rm F}\phi^{\rm q}({\bf r}) \sigma_x \psi^{\rm K'}({\bf r})
 +v_{\rm F}\bsigma \cdot (\hat{\bf p}+{\bf A}^{\rm q}({\bf r}))
 \psi^{\rm K}({\bf r}).
 \label{eq:psi_K'_T}
\end{align}
The equations in the last lines of Eqs.~(\ref{eq:psi_K_T})
and (\ref{eq:psi_K'_T}) are nothing but 
the second and first row of
Eq.~(\ref{eq:Weyl_KK'}), respectively.
Therefore, Eq.~(\ref{eq:Weyl_KK'_Aem}) 
(or Eq.~(\ref{eq:Weyl_KK'})) 
is symmetric under the time-reversal transformation 
(when ${\bf A}({\bf r})=0$).

When $\phi^{\rm q}({\bf r})=0$, it is useful to 
define an operation ${\cal S}$ that 
transforms the electron of ${\bf k}$ to that of ${\bf -k}$
within the same valley.
The operation ${\cal S}$ is defined by 
\begin{align}
 {\cal S}
 \begin{pmatrix}
  \psi^{\rm K}({\bf r}) \cr 
  \psi^{\rm K'}({\bf r})
 \end{pmatrix}
 =
 \begin{pmatrix}
  \sigma_y \psi^{\rm K}({\bf r})^* \cr 
  \sigma_y \psi^{\rm K'}({\bf r})^*
 \end{pmatrix}, \ {\rm where} \
 {\cal S}=
 \begin{pmatrix}
 I & 0 \cr 0 & I
 \end{pmatrix}
 \sigma_y K.
 \label{eq:strs}
\end{align}
This operation is referred to as 
effective time-reversal symmetry or 
special time-reversal symmetry because we obtain
${\cal S}^{-1}{\bf J}({\bf r}){\cal S}=
-{\bf J}({\bf r})$
for the special case of $\phi^{\rm q}({\bf r})=0$, 
${\bf A}^{\rm q}({\bf r})=0$, and ${\bf A}({\bf r})=0$.
By applying Eq.~(\ref{eq:strs}) to the 
wavefunction of Eq.~(\ref{eq:Weyl_KK'}), 
we get for the first row as
\begin{align}
 & 
 E \sigma_y \psi^{\rm K}({\bf r})^* =
 v_{\rm F}\bsigma \cdot 
 (\hat{\bf p}+{\bf A}^{\rm q}({\bf r})) \sigma_y \psi^{\rm K}({\bf r})^*
 + v_{\rm F}\phi^{\rm q}({\bf r}) \sigma_x \sigma_y \psi^{\rm K'}({\bf r})^*
 \nn \\
 & \stackrel{\sigma_y}{\to}
 E \psi^{\rm K}({\bf r})^* =
 v_{\rm F}\bsigma' \cdot 
 (\hat{\bf p}+{\bf A}^{\rm q}({\bf r})) \psi^{\rm K}({\bf r})^*
 - v_{\rm F}\phi^{\rm q}({\bf r}) \sigma_x \psi^{\rm K'}({\bf r})^*
 \nn \\
 & \stackrel{\rm c.c.}{\to}
 E \psi^{\rm K}({\bf r}) =
 v_{\rm F}\bsigma \cdot 
 (\hat{\bf p}-{\bf A}^{\rm q}({\bf r})) \psi^{\rm K}({\bf r})
 - v_{\rm F}\phi^{\rm q}({\bf r})^* \sigma_x \psi^{\rm K'}({\bf r})
 \label{eq:psi_K'_ST}
\end{align}
Here, from the first line to the second line,
we multiplied $\sigma_y$ to both sides and used
$\sigma_y \bsigma \sigma_y=\bsigma'$.
To get the last line,
we took the complex conjugation of the second line 
and used $(\bsigma')^*=-\bsigma$ and $\hat{\bf p}^* = -\hat{\bf p}$.
The special operation ${\cal S}$ becomes a symmetry 
when ${\bf A}^{\rm q}({\bf r})=0$.
Note that ${\bf A}^{\rm q}({\bf r})=0$ results in $\phi^{\rm q}({\bf r})=0$
because of Eq.~(\ref{eq:phi}).

We comment on the definition of the time-reversal operation
and the on-site interaction.
The transformation defined by Eq.~(\ref{eq:trs}) 
does not contain any Pauli spin matrix
since time-reversal symmetry has nothing to do with 
the pseudospin degree of freedom.
It is also noted that 
the pseudospin degree of freedom 
($\sigma_y$ in Eq.~(\ref{eq:strs})) 
is necessary to define 
the special symmetry of Eq.~(\ref{eq:strs}).
Even in the presence of the on-site deformation potential
of Eq.~(\ref{eq:phi_AB}),
the time-reversal symmetry of Eq.~(\ref{eq:trs}) is valid
since Eq.~(\ref{eq:trs}) does not contain the Pauli spin matrix.
However, the special symmetry of Eq.~(\ref{eq:strs}) is lost
when Eq.~(\ref{eq:phi_AB}) is not symmetric 
about the sublattice.~\cite{berry87}
Namely, when we take into account a non-vanishing $\sigma_z$ term 
of Eq.~(\ref{eq:onsite_psin}) for Eq.~(\ref{eq:psi_K'_ST}),
the symmetry of Eq.~(\ref{eq:strs}) is broken even when 
${\bf A}^{\rm q}({\bf r})=0$.


\section{Electron-Phonon Interaction}\label{sec:elph}

In this section, we apply the ${\bf A}^{\rm q}({\bf r})$
for phonon mode to describe the el-ph interaction.
There are six phonon modes in graphene:
in-plane longitudinal optical/acoustic mode (LO/LA),
in-plane tangential optical/acoustic mode (iTO/iTA), and
out-of-plane tangential optical/acoustic mode
(oTO/oTA).~\cite{saito98book} 
We consider long wavelength in-plane optical phonon modes;
the LO and iTO phonon modes near the $\Gamma$ point.
Using Eqs.~(\ref{eq:A}) and (\ref{eq:onsite_psin}),
we will derive ${\bf A}^{\rm q}({\bf r})$ and ${\cal H}_{\rm on}$
for the el-ph interaction of the LO/iTO modes.
The material in this section 
has been used to analyze the LO/iTO phonon frequency shift
in metallic SWNTs.~\cite{q1239}
The phonon frequency shift is observed 
as a function of the Fermi energy (the Kohn anomaly) 
in the Raman spectroscopy
experiments.~\cite{ferrari06,yan07,farhat07,nguyen07,wu07,das07}

\subsection{$\Gamma$ point optical phonon modes}\label{sec:G_op}

\begin{figure}[htbp]
 \begin{center}
  \includegraphics[scale=0.7]{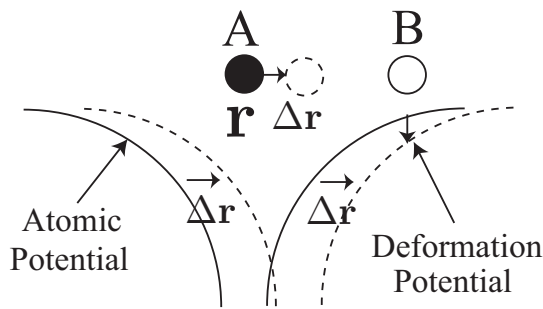}
 \end{center}
 \caption{
 The atomic potential of a carbon atom at ${\bf r}$
 is denoted by solid curves and the shifted potential 
 of the carbon atom at ${\bf r}+\Delta{\bf r}$ 
 is plotted by dashed curves.
 The deformation potential at the B atom is given by 
 $\Phi({\bf r}+\Delta {\bf r})- \Phi({\bf r}) = 
 \Delta{\bf r} \cdot \nabla \Phi({\bf r})+\cdots$.
 The deformation potential gives rise to a scattering
 of a $\pi$-electron from A-atom to the B-atom as 
 $\langle \Psi_{\rm B}({\bf r}+{\bf R}_a)|\nabla \Phi({\bf r})| \Psi_{\rm
 A}({\bf r}) \rangle \cdot ({\bf R}_a/a_{\rm cc})$.
 }
 \label{fig:defpot}
\end{figure}

The el-ph interaction originates from 
a change of the atomic potential
due to a vibration of a carbon atom.
In Fig.~\ref{fig:defpot}, we show a change of the atomic potential 
whereby $\delta \gamma^a_0({\bf r})$ is induced.
The atomic potential whose origin is located at ${\bf r}$
is denoted by solid curves and the shifted potential whose origin is
located at ${\bf r}+\Delta{\bf r}$ is plotted by dashed curves.
The deformation potential is given by 
$\Phi({\bf r}+\Delta {\bf r})- \Phi({\bf r}) = 
\Delta{\bf r} \cdot \nabla \Phi({\bf r})+\cdots$, which gives rise to 
matrix element between the $\pi$-electron at A-atom and  the
$\pi$-electron at B-atom as 
$\langle \Psi_{\rm B}({\bf r}+{\bf R}_a)|\nabla \Phi({\bf r})| \Psi_{\rm
A}({\bf r}) \rangle \cdot ({\bf R}_a/a_{\rm cc})$.
In this paper we denote this matrix element as $g_{\rm off}/\ell$,
that is, 
$\langle \Psi_{\rm B}({\bf r}+{\bf R}_a)|\nabla \Phi({\bf r})| \Psi_{\rm
A}({\bf r}) \rangle \cdot ({\bf R}_a/a_{\rm cc})=g_{\rm off}/\ell$.
According to density functional calculation 
by Porezag {\it et al.},~\cite{porezag95}
we have 
$g_{\rm off}/\ell =3.0$eV/\AA~\cite{jiang05prb,sasaki07local}
where $\ell\equiv 3a_{\rm cc}/2=2.13$\AA.

When $\Delta {\bf r}={\bf u}({\bf r})$
(${\bf u}({\bf r})$ is defined in Eq.~(\ref{eq:u_opt})), 
the change of the C-C bond-length
is given by ${\bf u}({\bf r}) \cdot ({\bf R}_a/a_{\rm cc})$.
Thus $\delta \gamma^a_0({\bf r})$ for the LO and TO modes is given by
\begin{align}
 \delta \gamma^a_0({\bf r})=
 \frac{g_{\rm off}}{\ell}
 {\bf u}({\bf r}) \cdot \frac{{\bf R}_a}{a_{\rm cc}}.
 \label{eq:dg_off}
\end{align}
By putting Eq.~(\ref{eq:dg_off}) into Eq.~(\ref{eq:A}),
we obtain
\begin{align}
 \begin{split}
  & v_{\rm F} A^{\rm q}_x({\bf r}) = \frac{g_{\rm off}}{\ell a_{\rm cc}}
  {\bf u}({\bf r}) \cdot \left(
  {\bf R}_1 -\frac{{\bf R}_2 + {\bf R}_3}{2} \right),
  \\ 
  & v_{\rm F} A^{\rm q}_y({\bf r}) = \frac{g_{\rm off}}{\ell a_{\rm cc}} 
  \frac{\sqrt{3}}{2} 
  {\bf u}({\bf r}) \cdot 
  \left( {\bf R}_2 - {\bf R}_3 \right).
 \end{split}
 \label{eq:A_opt_pre}
\end{align}
Because 
${\bf R}_1-({\bf R}_2+{\bf R}_3)/2=\ell {\bf e}_y$ 
and $\sqrt{3}/2({\bf R}_2-{\bf R}_3)=-\ell {\bf e}_x$
(see the caption of Fig.~\ref{fig:graphene}),
we can rewrite Eq.~(\ref{eq:A_opt_pre}) 
by ${\bf A}^{\rm q}({\bf r})$ as 
\begin{align}
 v_{\rm F} (A^{\rm q}_x({\bf r}),A^{\rm q}_y({\bf r})) 
 = \frac{g_{\rm off}}{a_{\rm cc}} (u_y({\bf r}),-u_x({\bf r})),
 \label{eq:optigauge}
\end{align}
where $u_x({\bf r}) \equiv {\bf u}({\bf r}) \cdot {\bf e}_x$
and 
$u_y({\bf r})\equiv {\bf u}({\bf r})\cdot {\bf e}_y$.~\cite{ishikawa06} 
For the on-site el-ph interaction of Eq.~(\ref{eq:onsite_psin}),
we can neglect the term which is proportional to $\sigma_0$
since the LO and TO optical phonon modes
satisfy ${\bf u}_{\rm A}({\bf r})= -{\bf u}_{\rm B}({\bf r})$.
Then, we get
\begin{align}
 {\cal H}_{\rm on} = 
 \frac{g_{\rm on}}{2} \sigma_z
 \nabla \cdot {\bf u}({\bf r}).
\end{align}
The resulting effective-mass Hamiltonian
for the K point and K' point become
\begin{align}
 \begin{split}
  & {\cal H}^{\rm K} = 
  v_{\rm F} \bsigma \cdot (\hat{\bf p}+{\bf A}^{\rm q}({\bf r}))+
  \frac{g_{\rm on}}{2} \sigma_z \nabla \cdot {\bf u}({\bf r}), \\
  & {\cal H}^{\rm K'} = 
  v_{\rm F} \bsigma' \cdot (\hat{\bf p}-{\bf A}^{\rm q}({\bf r}))+
  \frac{g_{\rm on}}{2} \sigma_z \nabla \cdot {\bf u}({\bf r}),
 \end{split}
 \label{eq:weyl_op_Gamma}
\end{align}
where ${\bf A}^{\rm q}({\bf r})$ is given by Eq.~(\ref{eq:optigauge}).
Putting Eq.~(\ref{eq:optigauge}) into Eq.~(\ref{eq:mag-field}),
we see in Eq.~(\ref{eq:weyl_op_Gamma}) that 
$\nabla \cdot {\bf u}({\bf r})$
is proportional to the deformation induced magnetic field
since 
\begin{align}
 \nabla \cdot {\bf u}({\bf r}) = 
 -\frac{v_{\rm F}a_{\rm cc}}{g_{\rm off}} B^{\rm q}_z({\bf r}).
\end{align}
Thus, if $B_z^{\rm q}({\bf r}) \ne 0$,
the energy spectrum for an electron
has a local energy gap around $E_{\rm F}$
because the terms proportional to $\sigma_z$ in the Hamiltonian
Eq.~(\ref{eq:weyl_op_Gamma})
work as time dependent mass term.~\footnote{
The mass term is defined by $m \sigma_z$ in the effective-mass
Hamiltonian where $m$ produces an energy gap in the energy spectrum.
}
However, the optical phonon mode does not generate a static 
(time independent) energy gap 
because ${\bf u}({\bf r})$ is oscillating 
as a function of time as ${\bf u}({\bf r})\cos (\omega t)$
so that the time average of $B_z({\bf r})$ field vanishes.
However, the energy gap can be oscillating
as a function of time when the vibration of atoms is coherent
in the space.

Next we consider an out-of-plane TO (oTO) phonon mode.
The oTO phonon eigenvector is pointing in the direction
of perpendicular to the 2D plane and 
does not give rise to a first order, in-plane bond-length change.
Thus, the off-site and on-site el-ph coupling
for the oTO mode are negligible 
as compared with those for in-plane LO and TO modes.

\subsubsection{Phonon softening for ${\bf q}= 0$ LO/iTO phonon mode}

\begin{figure}[htbp]
 \begin{center}
  \includegraphics[scale=0.6]{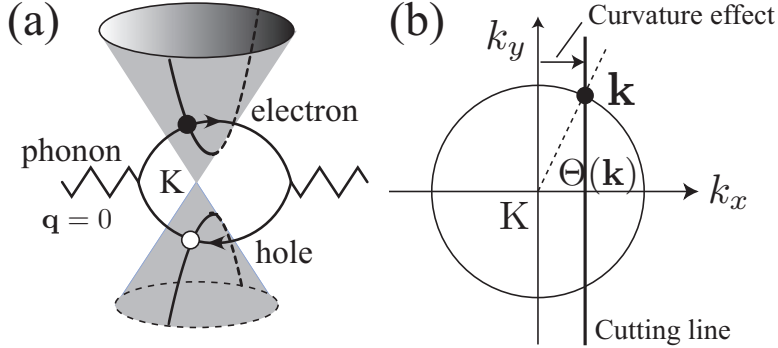}
 \end{center}
 \caption{ 
 (a) An electron-hole pair on the cutting line near the K point
 is excited by the $\Gamma$ point phonon mode (${\bf q}=0$).
 The intermediate electron-hole pair state contributes to the phonon
 frequency shift.
 (b)  The $k_x$ ($k_y$) axis corresponds to 
 the zigzag SWNT circumferential (axis)
 direction. The amplitude for an electron-hole pair creation
 (Eq.~(\ref{eq:coupling})) depends strongly on the relative position of
 the cutting line from the K point.
 If the cutting line crosses the K point, then
 the angle $\Theta({\bf k})$ ($\equiv \arctan(k_y/k_x)$)
 takes $\pi/2$ ($-\pi/2$) values for $k_y > 0$ ($k_y < 0$).
 In this case, the LO mode strongly couples to an electron-hole pair,
 while the TO mode is decoupled from the electron-hole pair.
 }
 \label{fig:cuttingline}
\end{figure}

For a uniform $(u_x,u_y)$ in the real space, that is, 
for the LO and TO modes 
with the phonon wave vector ${\bf q}={\bf 0}$, 
the el-ph interaction for the LO and TO phonon modes
is given only by ${\bf A}^{\rm q}$ in Eq.~(\ref{eq:weyl_op_Gamma}) since 
$\nabla \cdot {\bf u}({\bf r})=0$.
For the $\Gamma$ point optical phonon modes,
we consider a virtual electron-hole pair creation by the el-ph
interaction, which contributes to the phonon frequency shift
in second-order perturbation theory
(see Fig.~\ref{fig:cuttingline}(a)).
The phonon frequency shift as a function of the Fermi energy
has been observed in Raman spectroscopy
for graphene~\cite{ferrari06,yan07} and metallic
nanotubes.~\cite{farhat07,nguyen07,wu07,das07}

The el-ph matrix element 
for the electron-hole pair generation near the K point
is given from Eqs.~(\ref{eq:wfK}) and (\ref{eq:optigauge}) by
\begin{align}
 \langle {\rm eh}({\bf k})|{\cal H}_1^{\rm K}|\omega \rangle
 &= \int 
 (\psi^{\rm K}_{c,{\bf k}}({\bf r}))^* v_{\rm F} 
 \bsigma \cdot {\bf A}^{\rm op}
 \psi^{\rm K}_{v,{\bf k}}({\bf r}) d^2{\bf r}
 \nonumber \\
 &= \frac{g_{\rm off}}{2 a_{\rm cc}}
 \begin{pmatrix}
  e^{+i\frac{\Theta({\bf k})}{2}} \cr
  e^{-i\frac{\Theta({\bf k})}{2}}
 \end{pmatrix}^{t}
 \begin{pmatrix}
  0 & u_y + i u_x \cr
  u_y - i u_x & 0 
 \end{pmatrix} 
 \begin{pmatrix}
  e^{-i\frac{\Theta({\bf k})}{2}} \cr 
  -e^{+i\frac{\Theta({\bf k})}{2}}
 \end{pmatrix}.
 \label{eq:mat_elph}
\end{align}
Let us introduce the angle $\alpha$ 
between vector ${\bf u}=(u_x,u_y)$ 
and the $x$-axis,
then by putting $u_x=u\cos\alpha$ and $u_y=u\sin\alpha$
($u_x + iu_y=u e^{i\alpha}$) into Eq.~(\ref{eq:mat_elph})
we obtain 
\begin{align}
 \langle {\rm eh}({\bf k})|{\cal H}^{\rm K}_1|\omega_\alpha \rangle
 = -i g_{\rm off} \frac{u}{a_{\rm cc}} \cos (\Theta({\bf k})-\alpha).
 \label{eq:eh_G}
\end{align}
The el-ph matrix element for the electron-hole pair generation 
near the K' point, $\langle {\rm eh}({\bf k})|{\cal H}^{\rm
K'}_1|\omega_\alpha \rangle$, is the same as Eq.~(\ref{eq:eh_G}).
In Eq.~(\ref{eq:eh_G}),
$\alpha$ depends on the LO and TO modes in the case of SWNTs,
while $\alpha$ can not be defined uniquely in the case of flat graphene.
Let us consider the case of a zigzag SWNT.
Then, we denote $x$ ($y$) 
as a coordinate around (along) the axis
as shown in Fig.~\ref{fig:graphene}(b).
In this case,
${\bf u}=(u_x,0)$ ($\alpha=0$)
is assigned to the TO phonon mode while
${\bf u}=(0,u_y)$ ($\alpha=\pi/2$)
is assigned to the LO phonon mode.
By calculating Eq.~(\ref{eq:eh_G}) for the TO mode with 
$\alpha=0$ and for the LO mode with $\alpha=\pi/2$, 
we get 
\begin{align}
 \begin{split}
  &\langle {\rm eh}({\bf k})|{\cal H}_1^{\rm K}|\omega_{\rm TO} \rangle
  =-ig_{\rm off} \frac{u}{a_{\rm cc}} \cos \Theta({\bf k}), \\
  &\langle {\rm eh}({\bf k})|{\cal H}_1^{\rm K}|\omega_{\rm LO} \rangle
  =-ig_{\rm off} \frac{u}{a_{\rm cc}} \sin \Theta({\bf k}).
 \end{split}
 \label{eq:coupling}
\end{align}

For a ``metallic'' zigzag SWNT without the curvature
effect,~\cite{saito92prb,yang99}
we obtain $\Theta({\bf k})=\pm \pi/2$
for a metallic energy band with $k_x=0$
(see Fig.~\ref{fig:cuttingline}(b)).
Then, Eq.~(\ref{eq:coupling}) tells us that
only the LO mode gives rise to a non-vanishing el-ph matrix element
and the TO mode does not contribute to the electron-hole pair
creation.
The amplitude for an electron-hole pair creation 
depends strongly on the curvature effect
which shifts the relative position of the cutting line
from $k_x=0$ (see Fig.~\ref{fig:cuttingline}(b)).
When the curvature effect is taken into account,
$\cos \Theta({\bf k})=k_x/(k_x^2 + k_y^2)^{1/2}$
is nonzero due to $k_x\ne 0$.
Thus, the low energy electron-hole pair can couple to the TO phonon mode
and the matrix element becomes maximum when $k_y=0$.
On the other hand,
the high energy electron-hole pair still decouples to the TO phonon mode
since $\cos\Theta({\bf k}) \to 0$ for $|k_y|\gg |k_x|$.
This leads to a phonon frequency hardening 
for the TO phonon mode of a zigzag nanotube
when the Fermi energy is located near the Dirac point.~\cite{q1239}

The description of deformation induced gauge field 
for a lattice deformation (Eq.~(\ref{eq:A}))
is useful to show the appearance 
of the curvature-induced mini energy gap  
in metallic zigzag SWNTs.~\cite{kane97}
For a zigzag SWNT,
we have $\delta \gamma^1_0=0$ and 
$\delta \gamma^2_0=\delta \gamma^3_0\ne 0$ from the
rotational symmetry around the tube axis  (see
Fig.~\ref{fig:graphene}). 
Then, using Eqs.~(\ref{eq:A}),
we get that $A_x^{\rm q} = - \delta \gamma_0^2/v_{\rm F} $ 
and $A_y^{\rm q} = 0$ as the curvature effect.
The cutting line of $k_x=0$ for the metallic zigzag
nanotube is shifted by a finite constant value of $A_x^{\rm q}$ 
because of the AB effect 
for the deformation-induced gauge field ${\bf A}^{\rm q}({\bf r})$. 
This explains the appearance of the curvature-induced mini energy gap,
\begin{align}
 E_{\rm gap}= \frac{60({\rm meV}\cdot{\rm nm}^2)}{d_t^2},
\end{align}
where $d_t$ is the diameter of a metallic zigzag SWNT.

\subsubsection{${\bf q}\ne 0$, LO/iTO mode}

The description of the el-ph interaction 
as a gauge field ${\bf A}^{\rm q}({\bf r})$
can be extended to show the decoupling
between the TO mode with ${\bf q}\ne 0$ and the electrons
using a gauge symmetry argument.~\cite{q1239}
The TO phonon mode with ${\bf q}\ne 0$ does not change the
area of the hexagonal lattice but instead gives rise to a shear
deformation for the hexagonal unit cell of graphene.
Thus, the TO mode (${\bf u}_{\rm TO}({\bf r})$) satisfies
\begin{align}
 \nabla \cdot {\bf u}_{\rm TO}({\bf r}) = 0, \ \ \ \
 \nabla \times {\bf u}_{\rm TO}({\bf r}) \ne 0.
 \label{eq:TO}
\end{align}
Thus, the on-site deformation potential is zero
in Eq.~(\ref{eq:weyl_op_Gamma}).
Using Eqs.~(\ref{eq:optigauge}) and (\ref{eq:TO}), 
we see that 
the deformation-induced magnetic field,
$B^{\rm q}_z({\bf r})$, becomes zero instead 
the divergence of ${\bf A}^{\rm q}({\bf r})$ exists, which are
\begin{align}
 \begin{split}
  & (\nabla \times {\bf A}^{\rm q}({\bf r}))\cdot {\bf e}_z= 
  B^{\rm q}_z({\bf r})= 
  -\frac{g_{\rm off}}{v_{\rm F}a_{\rm cc}}
  \nabla \cdot {\bf u}_{\rm TO}({\bf r})=0,
  \\ 
  & \nabla \cdot {\bf A}^{\rm q}({\bf r})=
  \frac{g_{\rm off}}{v_{\rm F}a_{\rm cc}}
  (\nabla \times {\bf u}_{\rm TO}({\bf r}))\cdot {\bf e}_z
  \ne 0.
\end{split}
\end{align}
In this case, 
${\bf A}^{\rm q}({\bf r})$ can be represented by the gradient of 
a scalar function, $\Psi_a({\bf r})$, as
${\bf A}^{\rm q}({\bf r})=\nabla \Psi_a({\bf r})$.
Since we can choose a ${\bf A}^{\rm q}({\bf r})=0$ gauge
in Eq.~(\ref{eq:weyl_op_Gamma})
by a redefinition of the phase of the wavefunction
as $\psi^{\rm K}({\bf r}) \to \exp(-i\Psi_a({\bf r})/\hbar)
\psi^{\rm K}({\bf r})$ and 
$\psi^{\rm K'}({\bf r}) \to 
\exp(+i\Psi_a({\bf r})/\hbar) \psi^{\rm K'}({\bf r})$,~\cite{sasaki05}
the ${\bf A}^{\rm q}({\bf r})$ field in Eq.~(\ref{eq:weyl_op_Gamma})
disappears for the TO mode with ${\bf q}\ne 0$.
This explains why the TO mode with ${\bf q}\ne 0$ 
completely decouples from the electrons
and that only the TO mode with ${\bf q}= 0$
couples with electrons.
In this sense, the TO phonon mode at the $\Gamma$ point 
is anomalous since the el-ph interaction for the TO mode
can not be eliminated by a phase of the wavefunction. 
On the other hand, 
the LO phonon mode with ${\bf q}\ne 0$
changes the area of the hexagonal unit cell
while it does not give rise to a shear deformation.
Thus, the LO mode (${\bf u}_{\rm LO}({\bf r})$) satisfies
\begin{align}
 \nabla \cdot {\bf u}_{\rm LO}({\bf r}) \ne 0, \ \ \ \
 \nabla \times {\bf u}_{\rm LO}({\bf r}) = 0.
 \label{eq:LO}
\end{align}
Using Eqs.~(\ref{eq:optigauge}) and (\ref{eq:LO}), 
we see that the LO mode gives rise to a 
deformation-induced magnetic field as
\begin{align}
 B_z^{\rm q}({\bf r}) \ne 0, \ \ \
 \nabla \cdot {\bf A}^{\rm q}({\bf r})= 0.
\end{align}
Since a magnetic field changes the energy band structure
of electrons through the mass term,
the LO mode near the $\Gamma$ point (with ${\bf q}\ne 0$) 
is important for electronic energy spectrum.

\section{Edge States of graphene}\label{sec:edge}

\begin{figure}[htbp]
 \begin{center}
  \includegraphics[scale=0.6]{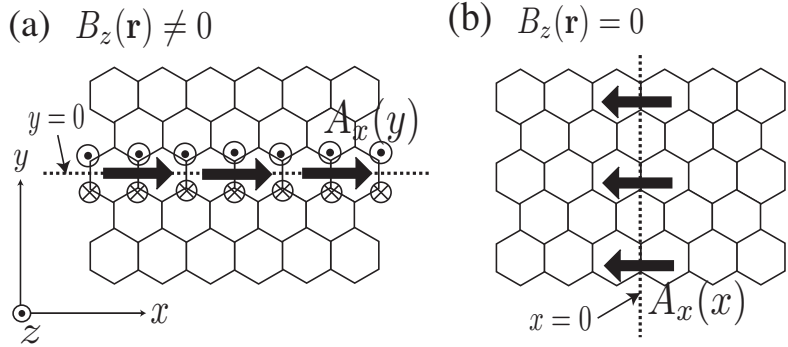}
 \end{center}
 \caption{
 Local deformation along a line and corresponding
 deformation-induced gauge field ${\bf A}^{\rm q}({\bf r})$
 are shown by arrows in (a) zigzag and (b) armchair edge.
 In (a), ${\bf A}^{\rm q}({\bf r})$ gives a finite deformation-induced 
 magnetic field $B_z^{\rm q}({\bf r})$ (flux) 
 illustrated by $\odot$ and $\otimes$ while no
 deformation-induced magnetic field is present in (b).} 
 \label{fig:graphene_edge}
\end{figure}

Finally, we give two examples of lattice deformation 
along a line in graphene
as shown in Fig.~\ref{fig:graphene_edge}(a) and (b) 
whose field $B^{\rm q}_z({\bf r})$ can polarize the pseudospin.
In Fig.~\ref{fig:graphene_edge}(a),
we modify the hopping integral at 
the dotted line of $y=0$,
so that $\delta \gamma_0^1(\mathbf{r})\ne 0$ at $y=0$ and 
$\delta \gamma_0^2(\mathbf{r}) = \delta \gamma_0^3(\mathbf{r})=0$.
From Eq.~(\ref{eq:A}), 
we obtain $v_F \mathbf{A}^{\rm q}(y) = (\delta \gamma_0^1(y),0)$.
In Fig.~\ref{fig:graphene_edge}(b),
the hopping integral along the dotted line at $x=0$
(precisely, $x=0_+$ or $x=0_-$) is changed as
$\delta \gamma_0^1(\mathbf{r})=0$ and 
$\delta \gamma_0^2(\mathbf{r}) = \delta \gamma_0^3(\mathbf{r}) \ne 0$ 
at $x=0$.
We obtain $v_F \mathbf{A}^{\rm q}(x) = (-\delta \gamma_0^2(x),0)$ 
for this case. 
Figure~\ref{fig:graphene_edge}(a) and (b)
correspond to the generation of the 
zigzag edge and the armchair edge 
in the limit of a strong perturbation, respectively,
if we take
$\delta \gamma_0^1= \gamma_0$ and 
$\delta \gamma_0^2=\delta \gamma_0^3= \gamma_0$
for the hopping integral at the two lines.
The behavior of the electronic structure
for the two cases are completely different from each other, which 
we show by calculating
the deformation-induced magnetic field.
In the case of Fig.~\ref{fig:graphene_edge}(a),
we have a finite deformation-induced magnetic field 
$B^{\rm q}_z({\bf r})\ne 0$ near the line at $y=0$.
The deformation-induced magnetic field 
$B^{\rm q}_z(y)$ is negative at $y < 0$ as illustrated by $\otimes$ 
and is positive at $y > 0$ as $\odot$.
On the other hand,
in the case of Fig.~\ref{fig:graphene_edge}(b),
the deformation-induced magnetic field is zero.
The gauge field which gives zero magnetic field can be removed from
Hamiltonian by a gauge transformation, which is discussed in the
previous section.
The deformation-induced magnetic field accounts for 
the presence of so-called edge states only at the zigzag edge, 
which is shown as below.
As we show in \S~\ref{sec:intro},
the edge state is a localized wavefunction near the zigzag edge
in which the pseudospin of the edge state is perfectly polarized.
This means that the amplitude of the wavefunction has a value only on
either A or B atoms.

\begin{figure}[htbp]
 \begin{center}
  \includegraphics[scale=0.6]{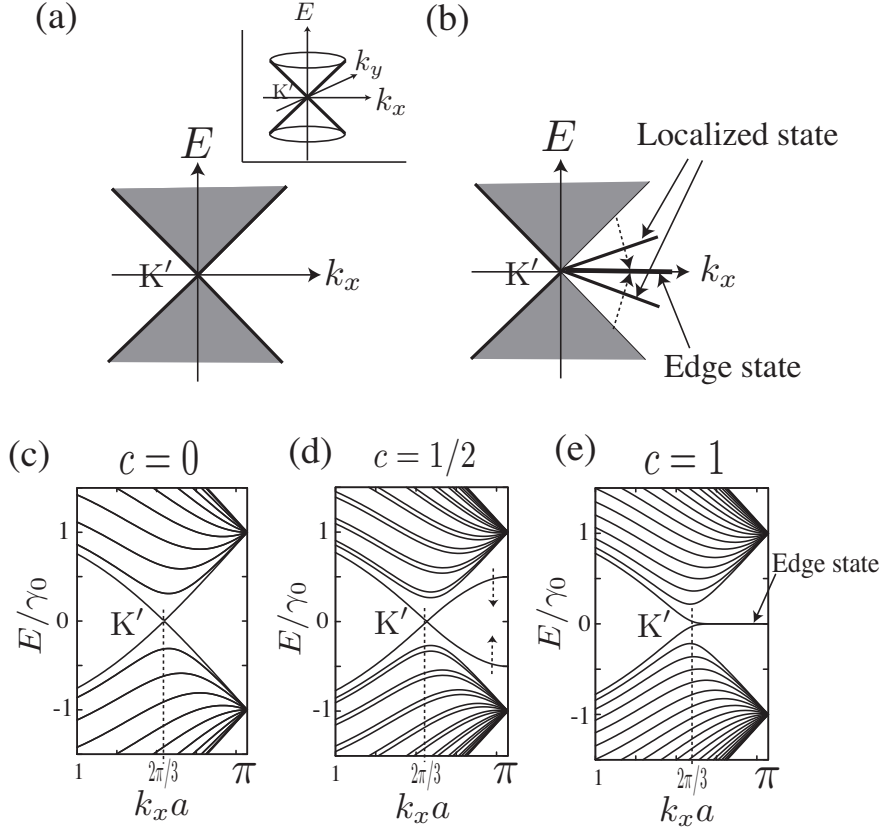}
 \end{center}
 \caption{
 (a) Band structure $(k_x,E)$ around the K' point of the effective-mass
 model without the deformation-induced gauge field and (b) that
 with the gauge field along the $x$-axis.
 The shaded regions in (a) and (b) represent 
 the spectrum of extended states for different $k_y$. 
 The (two-dimensional) dispersion relation around the K' point 
 for graphene without the deformation gives two cones 
 (representing the linear $k$ dispersion) whose
 apex is the K' point as is shown in the inset of (a).
 The band structure of (a) is a one-dimensional projection (onto $k_y$
 plane) of the linear $k$ dispersion.
 In (b), the energy dispersion for the localized state is represented by
 solid lines.
 When the bonds at $y=0$ become disconnected ($c=1$)
and zigzag edges appear,
 the energy eigenvalues of the localized state converge to zero ($E\to 0$)
 for any $k_x(>0)$ and forms a flat energy band of edge state.
 Using the TB model, we plot the band structure for an undeformed
 graphene in (c), for a zigzag ribbon in (e), and for a graphene with
 weakened hopping integral for the C-C bonds at $y=0$ in (d). 
 $k_x a$ in (c), (d) and (e) is the wave vector parallel to the $x$-axis
 and $k_x a=2\pi/3$ is the K' point.
 }
 \label{fig:cone}
\end{figure}

We will derive the edge states from the effective-mass model 
with the gauge field of Fig.~\ref{fig:graphene_edge}(a).~\cite{sasaki06jpsj}
Before going into the details, let us first outline the story.
We will show that there are localized pseudospin-polarized states 
in the energy spectrum
and the energy dispersion appears at $p_x > 0$ shown as 
the two solid lines in Fig.~\ref{fig:cone}(b).
The velocity for the energy dispersion
becomes small with increasing the gauge field 
and it becomes zero when the gauge field is sufficiently strong
which corresponds to the zigzag edge.
This result of the effective-mass model can
reproduce the result of the tight-binding (TB) lattice model
as shown in Fig.~\ref{fig:cone}(c),(d) and (e),~\cite{sasaki06jpsj} 
where the TB lattice model is defined by an adiabatic parameter $c$
as $\delta \gamma_0^1(y=0)=c \gamma_0$ in Eq.~(\ref{eq:H1}).
Here 
$c=0$ and $c=1$ correspond to no deformation and the zigzag edge,
respectively. 

\begin{figure}[htbp]
 \begin{center}
  \includegraphics[scale=0.45]{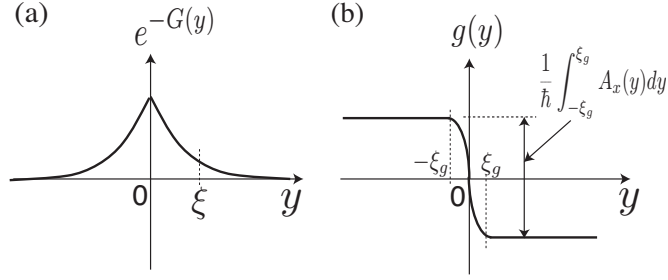}
 \end{center}
 \caption{(a) The amplitude of the wave function, $\exp(-G(y))$, of a 
 localized state whose localization length is $\xi$.
 (b) The pseudo-spin modulation part, $g(y)$.
 From Eq.~(\ref{eq:const-g(y)}), $g(y)$ is a constant 
 for $|y| \ge \xi_g$, and abruptly changes across the $x$-axis ($y=0$).  
 }
 \label{fig:Gandg}
\end{figure}

We assume that $\mathbf{A}^{\rm q}(\mathbf{r})$ 
of Fig.~\ref{fig:graphene_edge}(a) 
is quite localized within $|y|<\xi_g$,
that is,
$\mathbf{A}^{\rm q}(\mathbf{r})=(A_x^{\rm q}(y),0)$ and 
$A_x^{\rm q}(y)=0$ for $|y| \ge \xi_g$ in Eq.~(\ref{eq:HK'}),
where $\xi_g$ is a length of the order of lattice spacing.
We parameterize the localized energy eigenstate as
\begin{align}
 \psi^{\rm K'}_E(\mathbf{r}) = 
 N' \exp(ik_x x) e^{-G(y)}
 \begin{pmatrix}
  e^{+g(y)} \cr e^{-g(y)}
 \end{pmatrix},
 \label{eq:wfunc}
\end{align}
where $N'$ is a normalization constant.
The pseudospin polarization is 
represented by $g(y)$, and 
the wave vector $k_x$ is a good quantum number 
because of the translational symmetry along the $x$-axis.
In the direction of $y$,
we assume a localized nature of the wavefunction which was 
obtained by TB calculation.
Putting Eq.~(\ref{eq:wfunc}) to the energy eigenequation,
${\cal H}^{\rm K'}\psi^{\rm K'}_E(\mathbf{r})=E \psi^{\rm
K'}_E(\mathbf{r})$, we obtain
\begin{align}
 \begin{split}
  & p_x - A_x^{\rm q}(y) -\hbar \frac{d}{dy} \left(G(y)+g(y)\right) =
  - \frac{E}{v_F} e^{+2g(y)}, \\ 
  & p_x - A_x^{\rm q}(y) +\hbar \frac{d}{dy} \left(G(y)-g(y)) \right) = 
  - \frac{E}{v_F} e^{-2g(y)}.
 \end{split}
 \label{eq:w+g-d}
\end{align}
By summing and subtracting the both sides of Eq.~(\ref{eq:w+g-d}), 
we rewrite the energy eigenequation as 
\begin{align}
 \begin{split}
  & p_x -A_x^{\rm q}(y) - \hbar \frac{dg(y)}{dy} 
  = - \frac{E}{v_F} \cosh(2g(y)), \\
  & \hbar \frac{dG(y)}{dy} = \frac{E}{v_F} \sinh(2g(y)).
 \end{split} 
 \label{eq:weyl+g}
\end{align}
Since we are considering a localized solution around the $x$-axis,
we assume $G(y) \sim |y|/\xi$ 
where $\xi$ $(>0)$ is the localization length.
When $G(y) \sim |y|/\xi$, 
the solution of the second equation of Eq.~(\ref{eq:weyl+g})
is given by 
\begin{align}
 g(y) = \left\{
 \begin{array}{@{\,}ll}
  -\frac{1}{2} \sinh^{-1} \left( \frac{\hbar v_F}{\xi E}
  \right) & (y \le -\xi_g), \\
  +\frac{1}{2} \sinh^{-1} \left( \frac{\hbar v_F}{\xi E}
  \right) & (y \ge \xi_g). 
 \end{array} \right.
 \label{eq:const-g(y)}
\end{align}
The functions $G(y)$ and $g(y)$ 
are schematically shown in 
Fig.~\ref{fig:Gandg}(a) and (b), respectively.
The sign of $g(y)$ changes across the $x$-axis; 
the sign change of $g(y)$ means that the pseudospin
direction changes at the $x$-axis. 
The change of pseudospin is induced by the gauge field $A_x^{\rm q}(y)$.
To see this, we integrate the first equation of Eq.~(\ref{eq:weyl+g})
from $y=-\xi_g$ to $\xi_g$, and we get the following relation 
\begin{align}
 - \int_{-\xi_g}^{\xi_g} \frac{dg(y)}{dy} dy
 = \frac{1}{\hbar} \int_{-\xi_g}^{\xi_g} A_x^{\rm q}(y) dy.
 \label{eq:w-s-integ}
\end{align}
We have neglected other terms, since they are proportional to
$\xi_g$ and become zero in the limit of $\xi_g =0$.
By putting Eq.~(\ref{eq:const-g(y)}) to 
Eq.~(\ref{eq:w-s-integ}), we find
\begin{align}
 -\sinh^{-1} \left( \frac{\hbar v_F}{\xi E} \right) =
 \frac{1}{\hbar} \int_{-\xi_g}^{\xi_g} A_x^{\rm q}(y) dy.
 \label{eq:g-ene}
\end{align}
When the right-hand side of Eq.~(\ref{eq:g-ene}) is large,
we obtain from Eq.~(\ref{eq:const-g(y)}) that 
$g(y) \gg 0$ for $y \le -\xi_g$ and $g(y) \ll 0$ for $y \ge \xi_g$.
In this case, 
the localized state is a pseudospin-up state
$\psi^{\rm K}_{E}(\mathbf{r}) \propto {}^t (1,0)$
for $y \le -\xi_g$ and a pseudospin-down state
$\psi^{\rm K}_{E}(\mathbf{r}) \propto {}^t (0,1)$ for $y \ge \xi_g$.
Thus, a strong gauge field  at the $x$-axis makes  
pseudospin-polarized localized states.
Since the polarization of the pseudospin means that 
the wave function has amplitude only A (or B) atoms,
this result is consistent with the result by the TB model for the
edge state.~\cite{fujita96}
We understand that the edge state
is a pseudospin polarized and localized state
in the real space
which is induced by the deformation-induced magnetic field 
of the ${\bf A}^{\rm q}({\bf r})$ along the $x$-axis.

Let us now calculate $E$ and $\xi$.
To this end, 
we use the first equation of Eq.~(\ref{eq:weyl+g}) 
for $|y| \ge \xi_g$
and obtain 
\begin{align}
 \frac{E}{v_F}
 = \frac{-p_x}{\cosh \left( \displaystyle 
 \frac{1}{\hbar} \int_{-\xi_g}^{\xi_g} A_x^{\rm q}(y) dy \right)}.
 \label{eq:ene}
\end{align}
Moreover, using Eq.~(\ref{eq:g-ene}), we get
\begin{align}
 \frac{\hbar}{\xi}
 = p_x \tanh \left( \displaystyle 
 \frac{1}{\hbar} \int_{-\xi_g}^{\xi_g} A_x^{\rm q}(y) dy \right).  
 \label{eq:xi}
\end{align}
In addition to this localized state, 
there is another localized state for the same $p_x$ with 
the same $\xi$ but with an opposite sign of $E$.
This results from a particle-hole symmetry of the Hamiltonian;
$\sigma_z {\cal H}^{\rm K'} \sigma_z 
=-{\cal H}^{\rm K'}$.
By the particle-hole symmetry operation, 
the wave function is transformed as 
$\psi^{\rm K'}_{-E}(\mathbf{r}) = 
\sigma_z \psi^{\rm K'}_E(\mathbf{r})$.

The normalization condition of the wave function requires that 
$\xi$ should be positive, which 
restricts the value of $p_x$ to $p_x>0$ 
in Eq.~(\ref{eq:xi}).
In fact, when $A_x^{\rm q}(y)$ is positive, 
Eq.~(\ref{eq:xi}) means that the localized states appear only 
at $p_x>0$ around the K' point. 
This is the reason why 
the localized states appear in the energy spectrum
only in one side around the K' point 
as shown in Fig.~\ref{fig:cone}(b).
On the other hand, in the case of the K point 
the Hamiltonian is expressed by
Eq.~(\ref{eq:HK}).
Because of the different signs in front of 
$\mathbf{A}^{\rm q}(\mathbf{r})$
of Eqs.~(\ref{eq:HK}) and (\ref{eq:HK'}),
a similar argument gives $p_x < 0$ 
for the edge states around the K point.
Furthermore, 
when $(1/\hbar) \int_{-\xi_g}^{\xi_g} A_x^{\rm q}(y) dy \gg 0$, 
$E$ in Eq.(\ref{eq:ene}) becomes zero.
The zero energy eigenvalue between the K and K$'$ points in the band
structure corresponds to the flat energy band of the edge
state.~\cite{fujita96}
When $A_x^{\rm q}(y)$ is negative ($A_x^{\rm q}(y)\ll 0$),  
a flat energy band appears in the opposite side: $p_x < 0$
around the K' point and $p_x > 0$ around the K point.
This condition, $A_x^{\rm q}(y) \ll 0$, 
corresponds to the Klein's edges~\cite{klein94}
which are obtained by removing A or B atoms 
out of the zigzag edges having A or B atoms.
Calculated result by the TB model with the Klein's edges~\cite{klein94} 
is thus analytically explained by ${\bf A}^{\rm q}({\bf r})$.

There are also extended states in addition to the edge states.
The energy dispersion relation of extended states 
can be obtained as $(E/v_F)^2 = p_x^2 + p_y^2$, 
by setting $G(y) \sim -ip_y y/\hbar $ ($|y| \ge \xi_g$) in
Eqs.(\ref{eq:wfunc}) and (\ref{eq:weyl+g}) 
where $p_y$ is a real number.
The calculated energy bands are given by $|E|>v_F|p_x|$, 
shown as a shaded region in Fig.~\ref{fig:cone}(b). 
It also agrees well with the TB calculation 
shown in Fig.~\ref{fig:cone}(e).
Since $1-\tanh^2 x=1/\cosh^2 x$, 
we see that 
the energy dispersion relation of the localized state 
between $E$ and $\xi$ becomes
\begin{align}
 \left( \frac{E}{v_F} \right)^2 = p_x^2 - \left( \frac{\hbar}{\xi}
 \right)^2, \ \ (|y| \ge \xi_g),
 \label{eq:energydispersion}
\end{align}
which is the same as the linear dispersion relation 
$(E/v_F)^2 = p_x^2 + p_y^2$ 
if one replaces $p_y$ with $i(\hbar/\xi)$.
Thus One can then regard the localized state as a state with a
complex wavenumber.

We have shown that
three basic properties of the edge states, 
i.e., the pseudospin polarization,
the dependence on the momentum,
and the flat energy band,
obtained previously by the TB model,~\cite{fujita96} 
can be explained analytically 
in terms of the gauge field. 
In order to compare 
the present theory with the TB model quantitatively,
we have performed a TB calculation for the geometry of
Fig.~\ref{fig:graphene_edge}(a) with 
setting $\delta \gamma_0^1=c \gamma_0$. 
In Figs.~\ref{fig:cone}(c),~\ref{fig:cone}(d) and~\ref{fig:cone}(e), 
we plot the band structure for $c=0$, $c=1/2$, and $c=1$.
Comparing these figures with the results of the continuous model, one
can find an exact correspondence 
between the TB model and continuous model. 
Moreover, we analytically find~\cite{sasaki06jpsj}
\begin{align}
 \begin{split}
  & \frac{|E|}{v_F} = \frac{|p_x|}{\cosh (-\ln(1-c))} + 
  {\cal O}(l p_x^2/\hbar), \\
  & \frac{\hbar}{\xi} = p_x \tanh (- \ln (1-c)) + 
  {\cal O}(l p_x^2/\hbar),
 \end{split}
 \label{eq:lat-res-fin}
\end{align}
for localized states around the K' point.
Thus, by comparing Eq.~(\ref{eq:lat-res-fin}) with 
Eqs.~(\ref{eq:ene}) and (\ref{eq:xi}),
we conclude that the TB model and the continuous model 
agree with each other near the K' point ($p_x l/\hbar \ll 1$),
by the following relationship
\begin{align}
 \frac{1}{\hbar} \int_{-\xi_g}^{\xi_g} A_x^{\rm q}(y) dy 
 = - \ln (1-c).
 \label{eq:ans-gauge}
\end{align}
The right-hand side diverges when $c \to 1$, which 
reproduces the flat energy band ($E \to 0$) in Eq.~(\ref{eq:ene}) 
and gives $\xi/\hbar = p_x^{-1}$ in Eq.~(\ref{eq:xi}).
When $c \to 0$, we have $\frac{1}{\hbar} \int_{-\xi_g}^{\xi_g} A_x^{\rm
q}(y) dy \simeq c$, which confirms Eq.~(\ref{eq:A}) that is derived 
by assuming a weak perturbation for 
$\delta \gamma_0^a({\bf r})$.~\footnote{
It is interesting to note that
$B^{\rm q}_z(\mathbf{r})$ for the edge state corresponds to 
an enormous magnetic field $\sim 10^{5}$T at the zigzag
edge.~\cite{sasaki06jpsj,sasaki08jpsj} 
Thus a uniform external magnetic field has little effect on
the edge state, compared with the deformation.}

By considering the edge state using the effective-mass model, 
we found that the deformation-induced gauge field 
($\mathbf{A}^{\rm q}(\mathbf{r})$) 
and magnetic field ($B^{\rm q}_z(\mathbf{r})$)
explain basic properties of the edge state.
It is summarized as follows:\\
(1) The  gauge field can generate the edge state in energy spectrum,
depending on the gauge field direction.
Let $\mathbf{e}_\parallel$ the unit vector along the edge and 
$A^{\rm q}_\parallel(\mathbf{r}) \equiv 
\mathbf{A}^{\rm q}(\mathbf{r}) \cdot \mathbf{e}_\parallel$, 
the edge state appears if 
the gauge field has a component parallel to the edge: 
$A^{\rm q}_\parallel(\mathbf{r}) \ne 0$. 
This explains the presence (absence) of the edge state 
near the zigzag (armchair) edge. \\
(2) The edge states are pseudospin polarized states.
The direction of the pseudospin polarization is determined by the
direction or sign of $B^{\rm q}_z(\mathbf{r})$. \\  
(3) The direction of the gauge field, namely, 
$A^{\rm q}_\parallel(\mathbf{r}) > 0$ or $A^{\rm q}_\parallel(\mathbf{r}) < 0$,
is vital for the edge states, 
since it determines the energy dispersion and wave vectors 
which allow the edge states. \\
(4) The flat energy band of edge states at zero energy 
is obtained by the limit; 
$A^{\rm q}_\parallel(\mathbf{r}) \to \pm \infty$. \\

\section{Discussion and Summary}\label{sec:dis}

By formulating the effective-mass equation 
for a graphene with a lattice deformation, 
we have shown that the deformation can be represented 
by the deformation-induced gauge field, ${\bf A}^{\rm q}({\bf r})$.
We formulate the el-ph interactions 
and represents the shape of edge
by ${\bf A}^{\rm q}({\bf r})$.
The appearance of the gauge field in the effective-mass equation 
is reasonable because the Feynman diagram for the el-ph interaction
is basically the same as that for the electromagnetic interaction
of ${\bf A}({\bf r})$.
The only difference between the fields ${\bf A}^{\rm q}({\bf r})$
and ${\bf A}({\bf r})$ is that ${\bf A}^{\rm q}({\bf r})$
does not break time-reversal symmetry as a whole system
while ${\bf A}^{\rm q}({\bf r})$ breaks time-reversal symmetry 
locally in ${\bf k}$-space.
Thus, we think that a time-reversal symmetric gauge field 
is useful not only for graphene system 
but also other lattice systems that have an internal degree of freedom
like the pseudospin.

Here we would like to mention 
the extension of the effective-mass model for the edge states.
It is known that the Coulomb interaction 
makes the real spin (not pseudospin)
of the edge states polarized.~\cite{fujita96}
We considered the effective-mass model for the Hubbard model
with a mean field approximation,
and found that the Hubbard term appears as a mass term 
whose sign depends on the spin of the edge states.
The mass term shifts the energy of the edge states 
up or down relative to $E=0$ depending on the sign of the mass so that the
ground state exhibits the spin polarized state.
The gauge field and the mass term of the effective-mass model 
give rise to a parity anomaly phenomena.
We have shown that ferromagnetism of the edge states closely related to 
the parity anomaly.~\cite{sasaki08jpsj}


In summary,
the lattice deformation of graphene is modeled in the effective-mass
theory by the deformation-induced gauge field
which keeps the time-reversal symmetry 
but breaks the special symmetry defined within each valley.
The formalism of the gauge field and the gauge symmetry is
useful for understanding anomalous physical properties of graphene.

\section*{Acknowledgements}

The authors are grateful to Prof. S. Murakami 
(Tokyo Institute of Technology)
for his outstanding instruction for our collaborating works.
We also wish to thank Prof. Mildred Dresselhaus, Prof. Jing Kong, and
Dr. Hootan Farhat for sharing experimental data on Raman spectroscopy
of carbon nanotube.
R. S. acknowledges a Grant-in-Aid (Nos. 16076201 and 20241023) from
MEXT.


\end{document}